\documentclass[twocolumn,aps,showpacs,longbibliography]{revtex4-1}

\usepackage{graphicx}
\begin{document}

\title{Non-classicality and entanglement criteria for bipartite optical fields characterized by quadratic detectors II: Criteria based on probabilities}

\author{Jan Pe\v{r}ina Jr.}
\email{jan.perina.jr@upol.cz}
\author{}
\affiliation{Joint Laboratory of Optics of Palack\'{y} University and Institute of Physics of the Czech Academy of
Sciences, Faculty of Science, Palack\'{y} University, 17. listopadu 12, 77146 Olomouc, Czech Republic}
\author{V\'{a}clav Mich\'{a}lek}
\affiliation{Institute of Physics of the Czech Academy of
Sciences, Joint Laboratory of Optics of Palack\'{y} University and
Institute of Physics of CAS, 17. listopadu 50a, 772 07 Olomouc,
Czech Republic}
\author{Ond\v{r}ej Haderka}
\affiliation{Joint Laboratory of Optics of Palack\'{y} University and Institute of Physics of the Czech Academy of
Sciences, Faculty of Science, Palack\'{y} University, 17. listopadu 12, 77146 Olomouc, Czech Republic}

\begin{abstract}
Numerous non-classicality criteria based on the probabilities of experimental
photocount or theoretical photon-number distributions are derived using several
approaches. Relations among the derived criteria are revealed and the
fundamental criteria are identified. They are grouped into parametric systems
that allow the analysis of the non-classicality from different points of view
('local' non-classicality, pairwise character of photon correlations, etc.).
Considering their structure, the criteria may be divided into groups that
differ in the power to resolve the non-classicality. Quantification of the
non-classicality using the Lee non-classicality depth and the non-classicality
counting parameter is discussed. The used number of field's modes is identified
as an important parameter that may cause unexpected results. An appropriate
linear transformation of a photocount (photon-number) distribution into its
s-ordered form, needed for the determination of the non-classicality depth, is
derived. For comparison, the derived criteria are applied both to an
experimental photocount histogram of a twin beam and the reconstructed
photon-number distributions with different levels of the noise.
\end{abstract}

 \maketitle

\section{Introduction}

Non-classicality of optical
fields~\cite{Glauber1963,Sudarshan1963,Mandel1995,Vogel2001,Dodonov2003} and
their entanglement represent their most important properties useful in precise
(quantum) metrology (detection of gravitational waves), quantum and ghost
imaging \cite{Gatti2008,Brida2010a}, quantum communications (quantum key
distribution) and, in general, many quantum-information protocols needed in
quantum-information processing \cite{Nielsen2000}. Their identification and
even quantification in the case of real (experimental) optical fields represent
in general a very demanding task as the rigorous approaches are based upon the
properties of quasi-distributions in the phase space of the characterized
optical fields \cite{Vogel2001,Fedorov2014,Kenfack2004,Hill1997}. Such
quasi-distributions can be obtained by the homodyne detection combined with the
homodyne tomography~\cite{Leonhardt1997,Lvovsky2009}. This is feasible for
optical fields composed of one or at maximum several modes. For optical fields
composed of larger number of modes, like the usual twin beams are, this
approach is not applicable as such fields are properly characterized by their
quasi-distributions of (integrated) intensities, instead of amplitudes.
Moreover, the field intensity involves incoherent contributions from all
populated modes. Though the number of dimensions of such intensity
quasi-distributions (two for a twin beam) is incomparably smaller than the
number of dimensions of the original phase space, the reconstruction of
intensity quasi-distributions, especially of nonclassical and entangled
multi-mode optical fields, remains demanding owing to possible divergencies and
instabilities of the known reconstruction
methods~\cite{Perina1991,Haderka2005a,PerinaJr2013a}.

For this reason, the approach based on non-classicality (entanglement)
witnesses (identifiers) is widely used. Various inequalities may serve as such
witnesses. Such inequalities then represent non-classicality criteria (NCa, NC
when used in singular). We note that the NCa do not have to be applied directly
to an analyzed field, the analyzed field may be specifically modified, e.g., by
mixing with a defined coherent field at a beam splitter. This considerably
improves the ability of the NCa to reveal the entanglement
\cite{Kuhn2017,Arkhipov2018b}. We also note that, for simple systems, the
separability eigenvalue equations can be solved to unambiguously quantify the
entanglement in the analyzed field \cite{Sperling2019}. The fact that any NC
does not allow to identify the non-classicality of all non-classical states
represents a problem. It can be partly overcome when more NCa are considered
and their performance is mutually compared. For specific kinds of states,
suitable NCa can be identified. In the literature, the NCa based on intensity
moments are commonly used to reveal the non-classicality
\cite{Haderka2005a,Allevi2012,Allevi2013,Sperling2015,Harder2016,MaganaLoaiza2019}.
The NCa based on intensity moments were summarized in \cite{PerinaJr2017a}
where they were applied to twin beams. On the other hand, there also exist the
NCa based on the probabilities in photocount and photon-number distributions.
However, they have been applied only rarely
\cite{Klyshko1996,Hillery1985,Waks2004,Waks2006,Wakui2014,PerinaJr2017a} and no
systematic study of these NCa has been given so far. The reason may be that the
NCa with intensity moments allow for straightforward non-classicality
quantification by using the concept of the Lee non-classicality depth
(NCD)~\cite{Lee1991} and the accompanying well-known formula for the
transformation of intensity moments between different field orderings. In this
contribution, we give systematic analysis of such NCa based on the
probabilities. We also rewrite the concept of the Lee non-classicality depth
into the language of probabilities, which allows for quantification of the
non-classicality indicated by these NCa.

What benefit can be expected from the NCa for probabilities? First, we will
have more NCa at our disposal and so our chance to reveal the non-classicality
of any state will be greater. Here, we recall that these NCa operate in the
space of photocount (photon-number) distributions that is in certain sense
complementary to that of intensity (or photon-number) moments. The NCa based on
probabilities may be more sensitive than those for intensity moments for
various kinds of states (e.g., the non-classicality of a photon-number
distribution with only nonzero probabilities for even photon numbers obtained
from collinear parametric down-conversion is identified by the NC based on
probabilities \cite{Hillery1985} while the NCa for intensity moments fail).
Second, the NCa for probabilities may identify the non-classicality in specific
regions of the photocount or photon-number distributions. For example, the
central part, or the tails, of a distribution may be responsible for the field
non-classical character. They may also monitor specific features of the
analyzed field, e.g., the pairwise character of twin beams. Third, there are
different rules for the experimental errors of probabilities and intensity
moments. This favors the NCa for probabilities under specific conditions. For
example, the NCa for probabilities are more precise for the fields with low
mean photon numbers. On the other hand, when more intense fields are analyzed,
the determination of intensity moments that uses all elements of the
photon-number distribution involves averaging of the experimental errors of
individual probabilities which results in precisely determined lower-order
intensity moments.

The derivation of NCa for probabilities is based upon the Mandel detection
formula~\cite{Perina1991,Mandel1995} that provides a 'bridge' between the NCa
for probabilities and those for intensity moments. There is one to one
correspondence between both types of the NCa. However, and importantly, the
commonly used NCa for intensities involve only low-order intensity moments. The
reason is that the experimental error of an intensity moment increases with its
order. Taking into account the number of measurement repetitions only
lower-order intensity moments are usually available with sufficiently low
experimental errors. On the other hand, no such limitation occurs for
probabilities of photocount and photon-number distributions. This means that,
in our analysis, we need to derive and analyze the NCa for intensity moments of
arbitrarily high orders. So, in our contribution, we again consider the methods
summarized in Ref.~\cite{PerinaJr2017a}, but in their general formulation. We
find general relations and implications among the NCa derived by different
methods. This allows us to identify the fundamental general NCa useful in the
analysis of non-classicality of photocount and photon-number distributions.

We note that we apply the Mandel detection formula for an ideal detector. This
means that the derived NCa can be applied either to an experimental photocount
histogram (distribution of electrons in the detector) or the photon-number
distribution (of the field before detection) whatever their mutual relation is.
There also exist the NCa that use the photocount moments and give directly the
information about the non-classicality of the measured field. They either
somehow encompass the detector parameters \cite{Sperling2012a} or they are
'device independent' \cite{Sperling2017}. Such NCa are appealing for
application \cite{Bartley2013}.

We apply the derived NCa to an experimental twin beam that has been already
addressed in~\cite{PerinaJr2017a} from the point of view of the NCa for
intensity moments. There are two reasons. The NCa of both types can be divided
into groups with characteristic features. Our results reveal that the NCa of
both types from the corresponding groups exhibit comparable performance in
quantifying the non-classicality of twin beams. This poses the question about
the behavior of both types of the NCa for other types of optical fields.
Second, the determination of the Lee NCD for the NCa for intensity moments is
standardly done assuming one effective field mode (for a twin beam, one mode in
the signal beam, one in the idler beam). Here, we show that the number of modes
in an optical field is an important parameter in the determination of the NCD
that should be taken into account. Otherwise, non-physical results may occur,
especially in the case of highly nonclassical fields.

The NCa for probabilities also need different approach in their application
compared to those for intensity moments. This is related to the fact that
whereas we typically have at our disposal only a small number of intensity
moments, we usually have a greater number of probabilities, either in
photocount or photon-number distributions. Whereas we usually apply only couple
of suitable NCa for intensity moments chosen according to the type of the
state, we need to analyze greater number of the NCa for probabilities. This
needs suitable parameterizations. We suggest and develop some parameterizations
(e.g., using the majorization theory) and test them on the experimental twin
beam.

The paper is organized as follows. In Sec.~II, mapping between the NCa for
intensity moments and NCa for probabilities is discussed. Then the NCa for
intensity moments of arbitrary orders are derived and then transformed into
those for probabilities. The method using classically nonnegative polynomials
of intensities, the majorization theory, the Cauchy--Schwarz inequality and the
matrix method are in turn applied. Mutual general relations among the NCa
originating in different methods are revealed. Quantification of the
non-classicality identified by the NCa is addressed in Sec.~III including the
extension of the method of the Lee NCD to the domain of probability
distributions and multi-mode fields. In Sec.~IV, the performance of the NCa for
probabilities is compared with that of the NCa for intensity moments using the
experimental histogram of a twin beam analyzed in \cite{PerinaJr2017a}. Several
parameterizations of the NCa for probabilities are applied to the photon-number
distribution of the twin beam in Sec.~V. Two reconstructed photon-number
distributions originating in the maximum-likelihood reconstruction and the
Gaussian best fit are analyzed. The applied NCa for probabilities reveal,
together with the non-classicality and its location in the distributions,
different levels of the noise present in the reconstructed fields. Sec.~VI
brings conclusions. In Appendix A, the NCa for probabilities are explicitly
written for low photon numbers, in tight parallel to the NCa for intensity
moments addressed in~\cite{PerinaJr2017a}. They are useful for weak optical
fields at the single-photon level.

\section{Non-classicality criteria for probabilities}

The derivation of NCa for probabilities suggested by Klyshko~\cite{Klyshko1996}
exploits the Mandel detection formula~\cite{Perina1991,Mandel1995} that gives
the probability $ p(n) $ of detecting $ n $ photons in terms of the
quasi-distribution $ P_{\cal N} (W) $ of integrated intensity $ W $ related to
the normal ordering of field operators:
\begin{equation}  % 1
 p(n) = \frac{1}{n!} \int_{0}^{\infty} dW\, W^n \exp(-W) P_{\cal N}(W) .
\label{1}
\end{equation}
This formula can be rearranged into the form
\begin{equation}  % 2
 \frac{n! p(n)}{p(0)} = \frac{ \int_{0}^{\infty} dW\, W^n \exp(-W) P_{\cal N}(W)}{
  \int_{0}^{\infty} dW\, \exp(-W) P_{\cal N}(W) } \equiv \langle\langle W^n\rangle\rangle_{\rm mod}
\label{2}
\end{equation}
in which the r.h.s. gives the intensity moments of a certain
properly-normalized quasi-distribution. When bipartite optical fields
characterized by a quasi-distribution $ P_{{\rm si},\cal N}(W_{\rm s},W_{\rm
i}) $ of integrated intensities $ W_{\rm s} $ and $ W_{\rm i} $ are considered,
we arrive at the two-dimensional variant of Eq.~(\ref{2}):
\begin{equation}  % 3
 \frac{n_{\rm s}! n_{\rm i}! p_{\rm si}(n_{\rm s},n_{\rm i})}{p_{\rm si}(0,0)} =
  \langle\langle W_{\rm s}^{n_{\rm s}}
 W_{\rm i}^{n_{\rm i}} \rangle\rangle_{\rm mod}
\label{3}
\end{equation}
The relation (\ref{3}) then allows us to derive an NC for probabilities from
any NC written in intensity moments~\cite{PerinaJr2017a}. Examples of the
application of this approach are found in
Refs.~\cite{Waks2004,Waks2006,Wakui2014,PerinaJr2017a}.

The NCa for intensity moments are usually derived using the following four
methods: 1) the method based on nonnegative polynomials of intensities (the
corresponding NCa are denoted as $ E $ and $ \bar{E} $ below; symbols with bar
are reserved for the NCa for probabilities), 2) the Cauchy--Schwarz inequality
($C $, $ \bar{C}$), 3) the matrix method ($M $, $ \bar{M}$) and 4) various
variations of the majorization theory ($D $, $ \bar{D}$; $T $, $ \bar{T}$ in
Appendix~A). These methods were discussed in~\cite{PerinaJr2017a} from the
point of view of low-order intensity moments. Here, we utilize them again, in
their general formulation. We also derive mutual relations among the NCa
derived by different methods and identify the fundamental NCa. Using the
formula (\ref{3}) we derive the corresponding NCa for probabilities and write
them in suitable forms. The above mentioned methods are in turn discussed in
the following subsections.

\subsection{Non-classicality criteria based on nonnegative polynomials of intensities}

We consider the following two types of nonnegative polynomials of intensities
that classically give nonnegative mean values \cite{Lee1998,PerinaJr2019}. They
provide us the following two systems of NCa parameterized by three and four
indices ($ k_{\rm s}, k_{\rm i}, l_{\rm s}, l_{\rm i} \ge 0 $, $ l\ge 1 $):
\begin{eqnarray}  % 4-5
 & E_{k_{\rm s}k_{\rm i}l} = \langle W_{\rm s}^{k_{\rm s}} W_{\rm i}^{k_{\rm i}} (W_{\rm s}-W_{\rm i})^{2l}\rangle <0,&
\label{4} \\
 & E_{k_{\rm s}k_{\rm i}l_{\rm s}l_{\rm i}} = \langle W_{\rm s}^{k_{\rm s}} W_{\rm i}^{k_{\rm i}} (W_{\rm s}-\langle W_{\rm s}\rangle)^{2l_{\rm s}}
   (W_{\rm i}-\langle W_{\rm i}\rangle)^{2l_{\rm i}} \rangle <0.& \nonumber \\
 & &
\label{5}
\end{eqnarray}

The NCa $ E_{k_{\rm s}k_{\rm i}l} $ considered for $ l=1 $ and $ l=2 $ give us
two NCa for probabilities useful, e.g., in the analysis of twin beams [$
p(n_{\rm s},n_{\rm i}) \equiv p_{\rm si}(n_{\rm s},n_{\rm i}) $]:
\begin{eqnarray}   % 6-7
 & \bar{E}_{k_{\rm s}k_{\rm i}1} = \frac{k_{\rm s}+1}{k_{\rm i}}p(k_{\rm s}+1,k_{\rm i}-1) +
  \frac{k_{\rm i}+1}{k_{\rm s}}p(k_{\rm s}-1,k_{\rm i}+1)& \nonumber \\
  & - 2p(k_{\rm s},k_{\rm i}) <0, \hspace{2mm} k_{\rm s},k_{\rm i}\ge 1, &  \label{6} \\
 & \bar{E}_{k_{\rm s}k_{\rm i}2} = \frac{(k_{\rm s}+2)(k_{\rm s}+1)}{k_{\rm i}(k_{\rm i}-1)}p(k_{\rm s}+2,k_{\rm i}-2)
  + 6p(k_{\rm s},k_{\rm i})& \nonumber \\
  & + \frac{(k_{\rm i}+2)(k_{\rm i}+1)}{k_{\rm s}(k_{\rm s}-1)}p(k_{\rm s}-2,k_{\rm i}+2)
  - 4\frac{k_{\rm s}+1}{k_{\rm i}}p(k_{\rm s}+1,k_{\rm i}-1) & \nonumber \\
  & - 4\frac{k_{\rm i}+1}{k_{\rm s}}p(k_{\rm s}-1,k_{\rm i}+1) <0, \hspace{2mm} k_{\rm s},k_{\rm i}\ge 2. &  \label{7}
\end{eqnarray}
Similarly, the NCa $ E_{k_{\rm s}k_{\rm i}l_{\rm s}l_{\rm i}} $ for the lowest numbers $ l_{\rm s} $ and $ l_{\rm i} $
result in the NCa:
\begin{eqnarray}   % 8-9
 & \bar{E}_{k_{\rm s}k_{\rm i}10} = (k_{\rm s}+1)p(k_{\rm s}+1,k_{\rm i})p^2(0,0)
  + \frac{1}{ k_{\rm s} } p(k_{\rm s}-1,k_{\rm i}) &\nonumber \\
  & \times p^2(1,0) - 2p(k_{\rm s},k_{\rm i})p(1,0)p(0,0) <0, \hspace{2mm} k_{\rm s}\ge 1, k_{\rm i}\ge 0,& \nonumber \\
  & &  \label{8} \\
 & \bar{E}_{k_{\rm s}k_{\rm i}01} = (k_{\rm i}+1)p(k_{\rm s},k_{\rm i}+1)p^2(0,0)
  + \frac{1}{k_{\rm i}} p(k_{\rm s},k_{\rm i}-1) & \nonumber \\
  & \times p^2(0,1) - 2p(k_{\rm s},k_{\rm i})p(0,1)p(0,0) <0, \hspace{2mm} k_{\rm s}\ge 0, k_{\rm i}\ge 1.& \nonumber \\
  & &  \label{9}
\end{eqnarray}

\subsection{Non-classicality criteria based on the Cauchy--Schwarz inequality}

The Cauchy-Schwarz inequality $ |\int f g dP|^2 \le \int f^2 dP \int g^2 dP $
written for arbitrary functions $ f $ and $ g $ and classical probability
distribution $ P $, when applied with $ f = W_{\rm s}^{l_{\rm s}/2} W_{\rm
i}^{l_{\rm i}/2} W_{\rm s}^{k_{\rm s}} W_{\rm i}^{k'_{\rm i}} $, $ g = W_{\rm
s}^{l_{\rm s}/2} W_{\rm i}^{l_{\rm i}/2} W_{\rm s}^{k'_{\rm s}} W_{\rm
i}^{k_{\rm i}} $ and quasi-distribution $ P(W_{\rm s},W_{\rm i}) $ results in
the following classical inequalities ($  l_{\rm s}, l_{\rm i}, k_{\rm s},
k'_{\rm s}, k_{\rm i}, k'_{\rm i} \ge 0 $):
\begin{eqnarray}   % 10
 &\langle W_{\rm s}^{2k_{\rm s}+l_{\rm s}} W_{\rm i}^{2k'_{\rm i}+l_{\rm i}} \rangle
  \langle W_{\rm s}^{2k'_{\rm s} +l_{\rm s}} W_{\rm i}^{2k_{\rm i}+l_{\rm i}} \rangle & \nonumber\\
 & - \langle W_{\rm s}^{k_{\rm s}+k'_{\rm s}+l_{\rm s}} W_{\rm i}^{k_{\rm i}+k'_{\rm i}+l_{\rm i}} \rangle^2 \ge 0.&
\label{10}
\end{eqnarray}
Negation of the inequality (\ref{10}) with suitable relabelling of power
indices gives us the following NCa:
\begin{eqnarray}  % 11
 &C_{N}^{L} = \langle W^L \rangle \langle W^{2N-L}\rangle - \langle W^N\rangle^2 <0,& \nonumber\\
 & N \ge 0, 2N\ge L \ge 0.&
\label{11}
\end{eqnarray}
In Eq.~(\ref{11}) we apply the notation with vector indices $ N \equiv (n_{\rm
s},n_{\rm i}) $ in which $ W^N \equiv W_{\rm s}^{n_{\rm s}} W_{\rm i}^{n_{\rm
i}} $. The NCa $ C_{N}^{L} $ in Eq.~(\ref{11}) is then converted into the NCa
for probabilities:
\begin{equation}  % 12
 \bar{C}_{N}^{L} = \frac{(2N-L)! L!}{(N!)^2} p(L)p(2N-L) - p^2(N) <0;
\label{12}
\end{equation}
$ N! \equiv n_{\rm s}! n_{\rm i}! $ and $ p(N) \equiv p(n_{\rm s},n_{\rm i}) $.

\subsection{Non-classicality criteria reached by the matrix method}

The matrix method
\cite{Agarwal1992,Shchukin2005,Miranowicz2006,Vogel2008,Miranowicz2010,Sperling2015}
is based upon considering nonnegative quadratic forms of intensities
conveniently expressed in the matrix form. The obtained inequalities may even
be generalized using the Bochner theorem \cite{Richter2002,Ryl2015}. Using the
above vector-index notation for $ 2\times 2 $ matrices, we arrive at the
following NCa ($ L,N \ge 0 $):
\begin{eqnarray} % 13
 & {\rm det} \langle \left[ \begin{array}{cc} W^{2L} & W^{L+N} \\ W^{N+L} & W^{2N} \end{array} \right] \rangle =
  \langle W^{2L}\rangle \langle W^{2N}\rangle  - \langle W^{L+N} \rangle^2& \nonumber \\
  & < 0.&
\label{13}
\end{eqnarray}
However, these NCa are already involved in the NCa written in Eq.~(\ref{11}).

On the other hand, the $ 3\times 3 $ matrices of the form ($ K,L,N \ge 0 $)
\begin{eqnarray} % 14
 & {\rm det} \langle \left[ \begin{array}{ccc} W^{2K} & W^{K+L} & W^{K+N} \\ W^{L+K} & W^{2L} & W^{L+N} \\
  W^{N+K} & W^{N+L} & W^{2N} \end{array} \right] \rangle &
\label{14}
\end{eqnarray}
give rise to the following NCa for intensity moments:
\begin{eqnarray} % 15
 & M_{KLN} = \bigl\{ \langle W^{2K} \rangle [ \langle W^{2L}\rangle \langle W^{2N} \rangle - \langle W^{L+N} \rangle^2 ]  & \nonumber \\
 & + (KLN) \leftarrow (NKL) + (KLN) \leftarrow (LNK) \bigr\}  & \nonumber \\
 & + 2[\langle W^{K+L}\rangle \langle W^{K+N}\rangle \langle W^{L+N}\rangle - \langle W^{2K}\rangle \langle W^{2L}\rangle \langle W^{2N}\rangle] &
 \nonumber \\
 &  <0.  &
\label{15}
\end{eqnarray}
Symbol $ \leftarrow $ used in Eq.~(\ref{15}) replaces the terms that are
derived by the indicated change of indices from the explicitly written ones.
The corresponding NCa for probabilities are then derived in the form:
\begin{eqnarray} % 16
 & \bar{M}_{KLN} = \Bigl\{ p(2K)\Bigl[p(2L)p(2N)- \frac{(L+N)!^2}{(2L)!(2N)!} p^2(L+N)\Bigr] & \nonumber \\
 & + (KLN) \leftarrow (NKL) + (KLN) \leftarrow (LNK) \bigr\} & \nonumber \\
 & + 2\Bigl[ \frac{ (K+L)!(K+N)!(L+N)!}{(2K)!(2L)!(2N)!} p(K+L)p(K+N) & \nonumber \\
 & \times p(L+N) - p(2K)p(2L)p(2N)\bigr] <0.&
\label{16}
\end{eqnarray}

\subsection{Non-classicality criteria based on the majorization theory}

We first apply the majorization theory \cite{Marshall2010,Lee1990b} in the simplest case, i.e. with two variables ($
W_{\rm s} $, $ W_{\rm i} $) and quasi-distribution $ P( W_{\rm s} $,$ W_{\rm i}) $. It gives us the following NCa
\begin{equation} % 17
 \sum_{\{k+m,l-m\}} \langle  W_{\rm s}^{k+m}W_{\rm i}^{l-m}\rangle -
  \sum_{\{k,l\}} \langle  W_{\rm s}^{k}W_{\rm i}^{l}\rangle < 0
\label{17}
\end{equation}
for $ k \ge l $ and $ 1 \le m \le l $ because the orders $ \{k+m,l-m\} $ have
to majorize the orders $ \{k,l\} $ [ $ \{k+m,l-m\} \succ \{k,l\} $]. However,
detailed analysis of the NCa in Eq.~(\ref{17}) reveals that they can be
expressed as linear convex combinations of the NCa written in Eq.~(\ref{4})
and so they are redundant. As an example we consider in Eq.~(\ref{17}) the case
with $ k=l $ and $ m=2 $:
\begin{eqnarray} % 18
 & \langle  W_{\rm s}^{k+2}W_{\rm i}^{k-2}\rangle + \langle  W_{\rm s}^{k-2}W_{\rm i}^{k+2}\rangle
  - 2\langle  W_{\rm s}^k W_{\rm i}^k\rangle & \nonumber \\
 &  = E_{k+1,k-1,1} + E_{k-1,k+1,1} + 2E_{kk1}.&\label{18}
\end{eqnarray}
In general, it holds:
\begin{eqnarray} % 19
 & \langle  W_{\rm s}^{k+m}W_{\rm i}^{l-m}\rangle + \langle  W_{\rm s}^{l-m}W_{\rm i}^{k+m}\rangle
  - \langle  W_{\rm s}^k W_{\rm i}^l\rangle - \langle  W_{\rm s}^l W_{\rm i}^k\rangle & \nonumber \\
 & = E_{k+m-1,l-m+1,1} + 2E_{k+m-2,l-m+2,1} + \ldots + mE_{kl1} & \nonumber \\
 & + m E_{k-1,l+1,1} + m E_{k-2,l+2,1} + \ldots + mE_{l+1,k-1,1} & \nonumber \\
 & + mE_{lk1} + \ldots + 2E_{l-m+2,k+m-2,1} + E_{l-m+1,k+m-1,1}.& \nonumber \\
 & &\label{19}
\end{eqnarray}

Next, we consider the majorization inequalities for three variables and the
quasi-distributions $ P( W_{\rm s} $,$ W_{\rm i}) P_{a}(W'_a) $, $ a= {\rm s,i}
$, where $ P_{a} $ stands for the marginal quasi-distribution of $ P $.
Violation of the classical majorization inequality $ \{klm\} \succ \{k'l'm'\} $
gives us the NCa ($ a= {\rm s,i} $)
\begin{equation} % 20
 {^{a}D^{klm}_{k'l'm'}} =  f^a(k,l,m) - f^a(k',l',m') < 0
\label{20}
\end{equation}
expressed using the function $ f^a $:
\begin{eqnarray} % 21
 & f^a(k,l,m) = \{ (\langle W_{\rm s}^k W_{\rm i}^l \rangle + \langle W_{\rm s}^l W_{\rm i}^k \rangle )\langle W_a^m\rangle & \nonumber \\
 &  + (klm) \leftarrow (mkl) + (klm) \leftarrow (lmk) \} . &
\label{21}
\end{eqnarray}

The corresponding NCa for probabilities are expressed as [$ a= {\rm s,i} $; $
p_{\rm s}(m,n) \equiv p(m,n) $, $ p_{\rm i}(m,n) \equiv p(n,m) $]:
\begin{equation} % 22
 ^{a}\bar{D}^{klm}_{k'l'm'} = \bar{f}^a(k,l,m) - \bar{f}^a(k',l',m') < 0
\label{22}
\end{equation}
and
\begin{eqnarray} % 23
 & \bar{f}^a(k,l,m) = k!l!m!\{ [p(k,l) + p(l,k)]p_a(m,0) & \nonumber \\
 &+ (klm) \leftarrow (mkl) + (klm) \leftarrow (lmk) \} . &
\label{23}
\end{eqnarray}

From the NCa for probabilities written in Eq.~(\ref{22}), we consider the
following two three-parameter systems in further analysis:
\begin{eqnarray} % 24, 25
 & ^{a}\bar{\cal D}_{kl}^{1,m} = {^{a}\bar{D}^{k+m,k,l-m}_{kkl}} <0, \hspace{2mm} k\ge l \ge m\ge 1, \label{24}& \\
 & ^{a}\bar{\cal D}_{kl}^{2,m} = {^{a}\bar{D}^{k+m,k-m,l}_{kkl}} <0, \hspace{2mm} k-m\ge l \wedge k,m \ge 1 &\nonumber \\
 & \hspace{15mm} \wedge l \ge 0. \label{25}&
\end{eqnarray}

Also, another three-parameter system of the NCa for probabilities may be
defined by determining the minimum from the NCa based on 'moving one ball'
\cite{Marshall2010} in the majorization theory:
\begin{eqnarray}   % 26
 & ^{a}\bar{\cal D}_{klm} = \min\bigl[ ^{a}\bar{D}^{k+1,l-1,m}_{klm}, {^{a}\bar{D}^{k+1,l,m-1}_{klm}},
  {^{a}\bar{D}^{k,l+1,m-1}_{klm}} \bigr] & \nonumber \\
 & <0. & \label{26}
\end{eqnarray}
In the NCa on r.h.s. of Eq.~(\ref{26}), nonnegative $ k,l,m $ obey in turn the
inequalities $ k-1\ge l-1\ge m\ge 0 $, $ k\ge l\ge m\ge 1 $ and $ k-1\ge l\ge
m\ge 1 $. We note that the condition 'moving one ball' means that the
probabilities of two neighbor photon-numbers enter into the NCa.

From the point of view of revealing the non-classicality, the majorization
theory with four variables and quasi-distribution $ P( W_{\rm s},W_{\rm i})P(
W'_{\rm s},W'_{\rm i}) $ is the most promising. The NCa derived from the
classical majorization inequality $ \{klmn\} \succ \{k'l'm'n'\} $ take the form
\begin{equation} % 27
 D^{klmn}_{k'l'm'n'} = f(k,l,m,n) - f(k',l',m',n') < 0
\label{27}
\end{equation}
using the function $ f $:
\begin{eqnarray} % 28
 & f(k,l,m,n) = \{ (\langle W_{\rm s}^k W_{\rm i}^l \rangle + \langle W_{\rm s}^l W_{\rm i}^k \rangle )
  (\langle W_{\rm s}^m W_{\rm i}^n \rangle & \nonumber \\
 & + \langle W_{\rm s}^n W_{\rm i}^m \rangle ) + (klmn) \leftarrow (kmln) & \nonumber \\
 &  + (klmn) \leftarrow (knlm) \}. &
\label{28}
\end{eqnarray}

The corresponding NCa for probabilities obtained from Eq.~(\ref{27}) are
written as:
\begin{equation} % 29
 \bar{D}^{klmn}_{k'l'm'n'} = \bar{f}(k,l,m,n) - \bar{f}(k',l',m',n') < 0
\label{29}
\end{equation}
where
\begin{eqnarray} % 30
 & \bar{f}(k,l,m,n) = k!l!m!n!\{ [p(k,l) + p(l,k)][p(m,n) & \nonumber \\
 & + p(n,m)] + (klmn) \leftarrow (kmln) & \nonumber \\
 & + (klmn) \leftarrow (knlm) \} . &
\label{30}
\end{eqnarray}

From the NCa in Eq.~(\ref{29}) we consider the following three-parameter system
in the analysis of twin beams:
\begin{equation}   % 31
 \bar{\cal D}_{kl}^{3,m} = \bar{D}^{k+m,k,l,l-m}_{kkll} <0, \hspace{2mm} k\ge l\ge m \ge 1.
\label{31}
\end{equation}

Similarly as in the case with three variables, a useful NCa for probabilities
may be defined by taking the minimum from the NCa based on 'moving one ball' in
the majorization theory:
\begin{eqnarray}   % 32
 & \bar{\cal D}_{klmn} = \min\bigl[ \bar{D}^{k+1,l,m,n-1}_{klmn}, \bar{D}^{k+1,l,m-1,n}_{klmn},
  \bar{D}^{k+1,l-1,m,n}_{klmn}, & \nonumber \\
 & \bar{D}^{k,l+1,m,n-1}_{klmn}, \bar{D}^{k,l+1,m-1,n}_{klmn},
  \bar{D}^{k,l,m+1,n-1}_{klmn} \bigr]  <0. & \nonumber \\
 & & \label{32}
\end{eqnarray}
In the NCa on r.h.s. of Eq.~(\ref{32}), nonnegative $ k,l,m,n $ obey in turn
the inequalities $ k\ge l\ge m\ge n\ge 1 $, $ k-1\ge l-1\ge m-1\ge n\ge 0 $, $
k-1\ge l-1\ge m\ge n\ge0 $, $ k-1\ge l\ge m\ge n\ge 1 $, $ k-2\ge l-1\ge m-1
\ge n\ge 0 $, and $ k-1\ge l-1\ge m\ge n\ge 1 $.

The majorization theory suggests many other classical inequalities for testing
the non-classicality. They may be useful in the analysis of different kinds of
nonclassical states. As an example, to demonstrate versatility of the
majorization theory we derive the NCa using the inequality $ \{
\underbrace{k+m,l+n,0,0,\ldots }_{k+l+2} \} \succ \{ \underbrace{m,n,1,1,\ldots
}_{k+l+2} \} $ and quasi-distribution $ \prod_{j=1}^{(k+l)/2+1} P( W_{j,{\rm
s}},W_{j,{\rm i}}) $ ($ k,l \ge 0 $, $ m \ge n \ge 1 $):
\begin{eqnarray} % 33
 & {\cal D}_{kl}^{mn} = g(k+m,l+n) +\frac{k+l}{2} g(k+m,0) g(l+n,0) & \nonumber \\
 & - \bigl[ \frac{1}{2}g(m,n) g(1,1) + \frac{k+l}{2} g(m,1) g(n,1) \bigr] & \nonumber \\
 & \times \left[ \frac{1}{2}g(1,1)\right]^{(k+l)/2-1} <0,& \label{33}\\
 & g(k,l) = \langle W_{\rm s}^{k}W_{\rm i}^{l}\rangle + \langle W_{\rm s}^{l}W_{\rm i}^{k}\rangle .& \nonumber
\end{eqnarray}
The NCa in Eq.~(\ref{33}) when written in probabilities take the form:
\begin{eqnarray} % 34
 & \bar{\cal D}_{kl}^{mn} = \frac{(k+m)!(l+n)!}{m!n!}\bigl[ \bar{g}(k+m,l+n)p(0,0) & \nonumber \\
 & +\frac{k+l}{2} \bar{g}(k+m,0) \bar{g}(l+n,0) \bigr] p(0,0)^{(k+l)/2-1} & \nonumber \\
 & - \bigl[ \bar{g}(m,n)p(1,1) + \frac{k+l}{2} \bar{g}(m,1) \bar{g}(n,1) \bigr]p(1,1)^{(k+l)/2-1} & \nonumber \\
 & <0,& \label{34} \\
 & \bar{g}(k,l) = p(k,l) + p(l,k).& \nonumber
\end{eqnarray}

We note that the majorization theory has been found useful also in
identification of the non-classicality of single mode fields
\cite{Lee1990a,Verma2010,Arkhipov2016c,PerinaJr2017c}.

According to the analysis in the above subsections and taking into account
relations among the NCa, we have the following parametric systems of the NCa
for probabilities at disposal to investigate the non-classicality of photocount
and photon-number distributions. The consideration of nonnegative polynomials
gives us four two-parameter systems described in Eqs.~(\ref{6})---(\ref{9}). We
may apply the four-parameter system in Eq.~(\ref{12}) originating in the
Cauchy-Schwarz inequality or the six-parameter system in Eq.~(\ref{16}) derived
by the matrix method. We have also grouped the NCa from the majorization theory
into four three-parameter systems given in Eqs.~(\ref{24}), (\ref{25}),
(\ref{26}), and (\ref{31}) and two four-parameter systems described in
Eqs.~(\ref{32}) and (\ref{34}).

\section{Quantification of the non-classicality}

To convert the NCa into non-classicality quantifiers, the concept of NCD
introduced by Lee~\cite{Lee1991} is commonly applied using the intensity
moments and assuming the field composed of one effective mode ($ M=1 $, see
below). The intensity moments $ \langle W^k\rangle $, $ k=1,\ldots $, of the
analyzed field determined for the normal ordering of field operators are first
transformed into the $ s $-ordered moments $ \langle W^k\rangle_s $ along the
formula~\cite{Perina1991}:
\begin{eqnarray}  % 35
 \langle W^k\rangle_s = \frac{k!}{\Gamma(k+M)} \left(\frac{1-s}{2}\right)^k \left\langle L_k^{M-1}\left( \frac{2W}{s-1}
  \right) \right\rangle, \nonumber \\
\label{35}
\end{eqnarray}
where $ L_k^{M-1} $ are the Laguerre polynomials~\cite{Morse1953}, $ \Gamma $
denotes the Gamma-function and $ M $ gives the number of field modes. Then, the
$ s $-ordered intensity moments are inserted into the NCa and the value $
s_{\rm th} $ of ordering parameter that nullifies a given NC gives the
corresponding NCD $ \tau $ by the formula:
\begin{equation} % 36
 \tau = \frac{1-s_{\rm th} }{2} .
\label{36}
\end{equation}

We have to translate the transformation between different field orderings into
the language of photon-number distributions to allow its application in
relation to the above derived NCa for probabilities. We proceed as follows. We
define a photon-number distribution $ p_s(n) $ using the standard Mandel
detection formula in Eq.~(\ref{1}) in which, however, an $ s $-ordered
quasi-distribution $ P_s(W) $ of integrated intensity $ W $ is considered.
According to Ref.~\cite{Perina1991}, we have
\begin{eqnarray} % 37
 P_s(W)&=& \frac{1}{W} \exp\left(-\frac{2W}{1-s}\right) \left(\frac{2W}{1-s}\right)^M \nonumber \\
 & & \hspace{-12mm} \times \sum_{n=0}^{\infty}  \frac{n!p(n)}{[\Gamma(n+M)]^2}
  \left(\frac{s+1}{s-1}\right)^n L_n^{M-1}\left(\frac{4W}{1-s^2}\right) .
\label{37}
\end{eqnarray}
Substituting Eq.~(\ref{37}) into Eq.~(\ref{1}) the following linear relation
between the photon-number distributions $ p_s $ and $ p $ is obtained:
\begin{equation} % 38
 p_s(n) = \sum_{m=0}^{\infty} K_s(n,m;M) p(m),
\label{38}
\end{equation}
where
\begin{eqnarray} % 39
 K_s(n,m;M) &=& \left(\frac{2}{3-s}\right)^M \left(\frac{1+s}{1-s}\right)^m
  \left(\frac{1-s}{3-s}\right)^n  \nonumber \\
  & & \hspace{0mm} \times \sum_{l=0}^{m} (-1)^{m-l} \left(\begin{array}{c} m \\ l \end{array}\right)
  \left(\begin{array}{c} n+l+M-1\\ n \end{array}\right) \nonumber \\
  & & \times \left(\frac{4}{(1+s)(3-s)} \right)^l .
\label{39}
\end{eqnarray}
We note that the sum in Eq.~(\ref{39}) requires high numerical precision when
evaluated. For $ s=1 $, i.e. for the normal ordering, we have $ K_{s=1}(n,m) =
\delta_{nm} $ using the Kronecker symbol $ \delta_{nm} $. This can be verified
knowing that the function $ S(n,m,\alpha) $,
\begin{eqnarray} % 40
 S(n,m,\alpha) = 2^{m-n} \sum_{l=0}^{m} (-1)^{m-l}
  \left(\begin{array}{c} m \\ l \end{array}\right)
  \left(\begin{array}{c} n+l+\alpha\\ n \end{array}\right) \nonumber \\
 & &
\label{40}
\end{eqnarray}
equals zero for $ n<m $ and one for $ n=m $ independently on the value of a real
parameter $ \alpha \ge 0 $. Also, $ \sum_{n=0}^{\infty} K_s(n,m) = 1 $ which
can be checked using the Taylor expansion of function $ (1-x)^{-\alpha-1} $ at
$ x=0 $.

There exists an alternative to NCD --- a non-classicality counting parameter
(NCCP) $ \nu $ suggested in~\cite{PerinaJr2019}. Its definition is based upon
mixing the analyzed field with an additional $ M $-mode thermal field with the
varying amount of the noise. We have for this case
\begin{equation} % 41
 K_\nu(n,m;M) = p_{\rm M-R}(n-m;\nu,M)
\label{41}
\end{equation}
where the Mandel-Rice distribution $ p_{\rm M-R} $ for $ M $ modes with $ \nu $
mean thermal photons per mode is expressed as:
\begin{equation} % 42
 p_{\rm M-R}(n;\nu,M) = \frac{\Gamma(n+M)}{n!\, \Gamma(M)} \frac{\nu^n}{(1+\nu)^{n+M}} .
\label{42}
\end{equation}
Similarly as in the case of NCD, the non-classicality is quantified by the
maximal amount of the noise $ \nu $ that preserves the non-classicality of the
mixed field. We note that the definition of NCCP has clear physical motivation,
in contrast with the definition of the Lee NCD that is based upon a Gaussian
filter in the space of characteristic functions of phase-space
quasi-distributions~\cite{Perina1991}. However, the NCCP cannot be applied to
highly nonclassical fields~\cite{PerinaJr2019}.

\section{Application of non-classicality criteria to fields with low photocount numbers}

Here, we return to the original photocount histogram analyzed in
Ref.~\cite{PerinaJr2017a} from the point of view of the NCa for intensity
moments to extend the analysis to the NCa for probabilities. We show that the
NCa for intensity moments and probabilities can be divided into the
corresponding groups that differ in the structure and that exhibit different
performance in revealing the non-classicality. Moreover, we reveal the
important role of the number $ M $ of field modes ($ M=1 $ is standardly
assumed) in the determination of the Lee NCD. We also compare the performance
of two non-classicality quantifiers, the NCD and the NCCP, suitable for the NCa
for probabilities. In more detail, we analyze the experimental photocount
histogram $ f_{\rm si}(c_{\rm s},c_{\rm i}) $ that characterizes a twin beam
containing 8.8 photon pairs per single detection interval and little amount of
the noise. The twin beam originated in type-I down-conversion in a $ \beta
$-barium-borate crystal pumped by an ultra-short pulse. Detection efficiency of
the intensified CCD (iCCD) camera~\cite{PerinaJr2017a} used in the measurement
was around 23~\% and the mean photocount numbers $ \langle c_{\rm s}\rangle $
and $ \langle c_{\rm i}\rangle $ equaled 2.21 and 2.23, respectively. Details
about the experiment and the obtained experimental photocount histogram $
f_{\rm si}(c_{\rm s},c_{\rm i}) $ [see Fig.~2(a) in~\cite{PerinaJr2017a}] are
given in Ref.~\cite{PerinaJr2017a}. The analysis is based upon the NCa derived
in Sec.~II and written explicitly for low intensity moments in
Ref.~\cite{PerinaJr2017a} and for probabilities of detecting low photocount
numbers in Appendix~A.

\begin{figure} % figs. 1a, b
 \resizebox{.47\hsize}{!}{\includegraphics{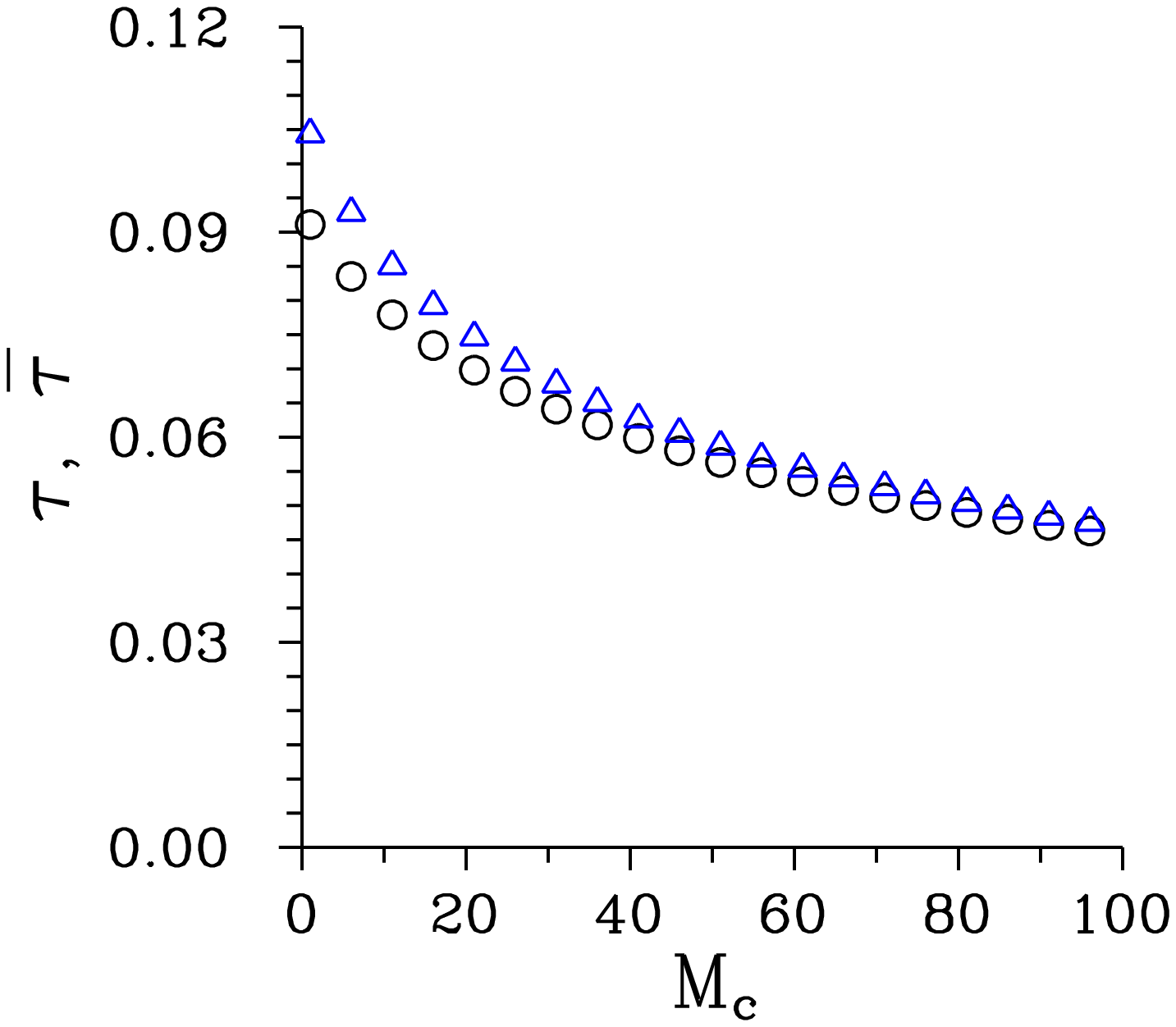}}  \hspace{1mm}
 \resizebox{.47\hsize}{!}{\includegraphics{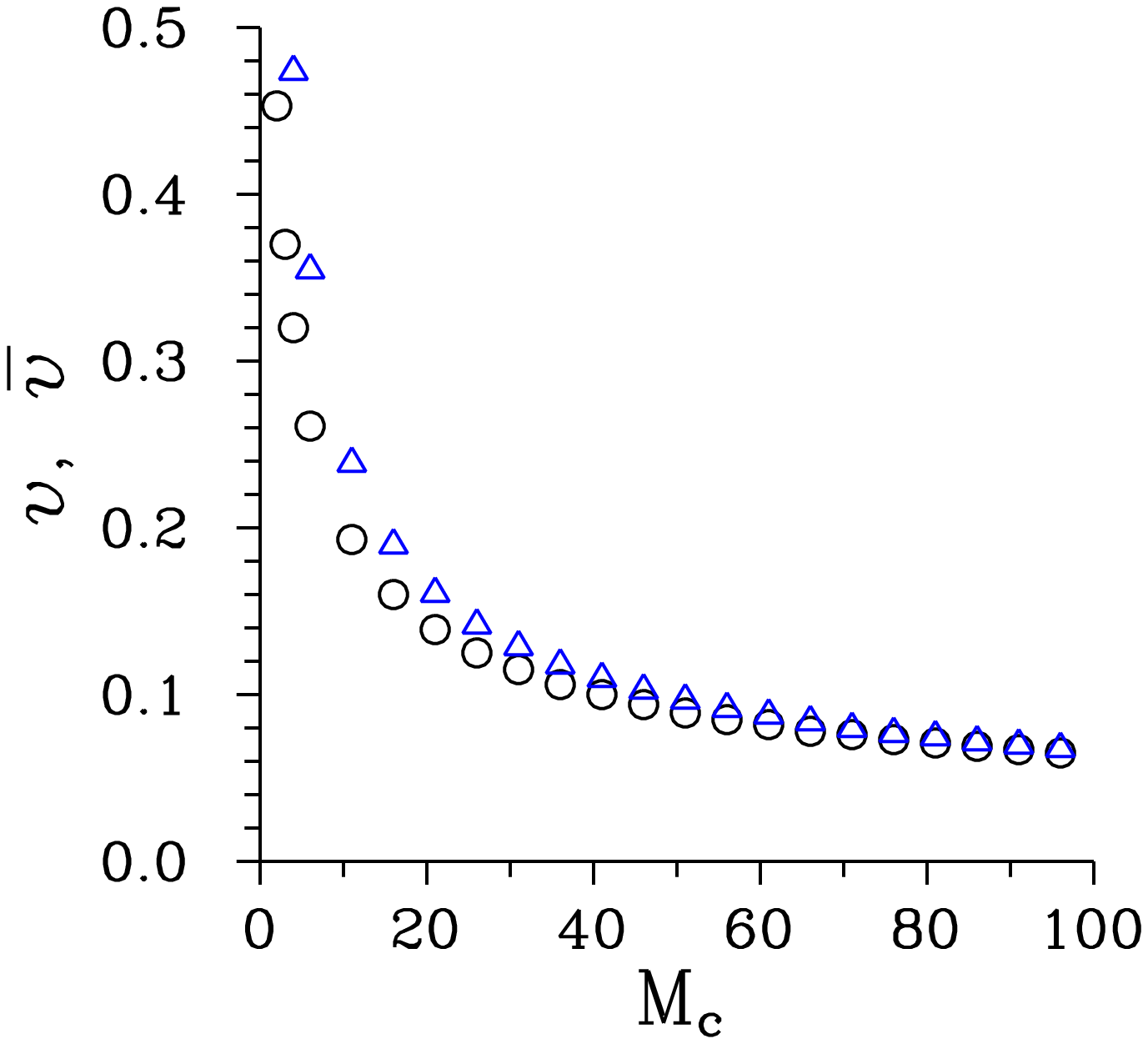}} \\
 \centerline{\small (a) \hspace{.4\hsize} (b)}
 \caption{(a) Non-classicality depths $ \tau $, $ \bar{\tau} $ and (b) non-classicality counting
  parameters $ \nu $, $ \bar{\nu} $ for NCa  $ E_{001} $, $ \bar{E}_{001} $ as they
  depend on number $ M_c $ of modes considering the NCa for intensity moments ($ \tau $, $ \nu $; black $ \circ $) and
  probabilities ($ \bar{\tau} $, $ \bar{\nu} $; blue $ \triangle $) and using the photocount histogram.
  Relative experimental errors are lower than 3~\%.}
 \label{fig1}
\end{figure}

We first address the role of the number $ M $ of field modes occurring in the
definitions of non-classicality quantifiers $ \tau $, $ \bar{\tau} $ and $ \nu
$, $ \bar{\nu} $. To do this, we chose the simplest NCa $ E_{001} $, $
\bar{E}_{001} $ that use up to the second-order intensity moments and
probabilities of detecting up to two photons [Eqs.~(\ref{6}), (\ref{A1})].
Their application to the experimental histogram $ f_{\rm si} $ reveals (see
Fig.~\ref{fig1}) that the greater the number $ M $ of modes the smaller the
values of NCDs $ \tau $, $ \bar{\tau} $ and NCCPs $ \nu $, $ \bar{\nu} $.
This behavior reflects the fact that the overall mean number of noisy photons
added to the field to conceal its non-classicality linearly increases with the
number $ M_c $ of modes ($ M_c\tau $, $ M_c\nu $). For practical application,
the number $ M $ of modes determined from the formula for multi-mode chaotic
field [$ M = \langle W\rangle^2/ \langle (\Delta W)^2\rangle $] may be a good
choice ($ M \approx 80$ for our data).

Then, we mutually compare the performance of the NCD and NCCP. According to the
graphs in Fig.~\ref{fig1}, fixing the number $ M_c $ of field modes the values
of NCCPs $ \nu $, $ \bar{\nu} $ are systematically greater than those of NCDs $
\tau $, $ \bar{\tau} $. Whereas the NCDs $ \tau $, $ \bar{\tau} $ for a single
mode Gaussian field have to be lower or equal to 1/2 (lower than one for a
general field), there is no upper bound for the NCCPs $ \nu $, $ \bar{\nu} $.
It has been discussed in Ref.~\cite{PerinaJr2019} that the NCCP cannot quantify
the non-classicality of highly nonclassical states. On the other hand, the
larger dynamical range observed for the NCCPs $ \nu $, $ \bar{\nu} $ in
Fig.~\ref{fig1} for small numbers $ M_c $ of modes may be advantageous. The
general applicability of the NCD, in contrast to the NCCP, is already
'embedded' in its definition. The comparison of both definitions in
Eqs.~(\ref{39}) and (\ref{41}) reveals the remarkable difference. The
physically-sound definition of NCCP $ \nu $ via the matrix $ K_\nu $ in
Eq.~(\ref{41}) describes 'shifting of the photon numbers of the field towards
greater values' due to the added noisy photons. On the other hand, the matrix $
K_s $ in Eq.~(\ref{39}) transforms a given photon number of the analyzed field
statistically into all possible (even lower) photon numbers. This is a
consequence of the fact that the ordering parameter $ s $ characterizes a
Gaussian filter for the characteristic function of the quasi-distribution of
amplitudes \cite{Perina1991}. We note that a numerical determination of the
matrix $ K_s $ is demanding. We also note that the NCa for multi-mode optical
fields were discussed in Refs.~\cite{Allevi2012,Allevi2012a}.

\begin{figure} % figs. 2a, b
 {\small (a)} \hspace{2mm}\resizebox{.8\hsize}{!}{\includegraphics{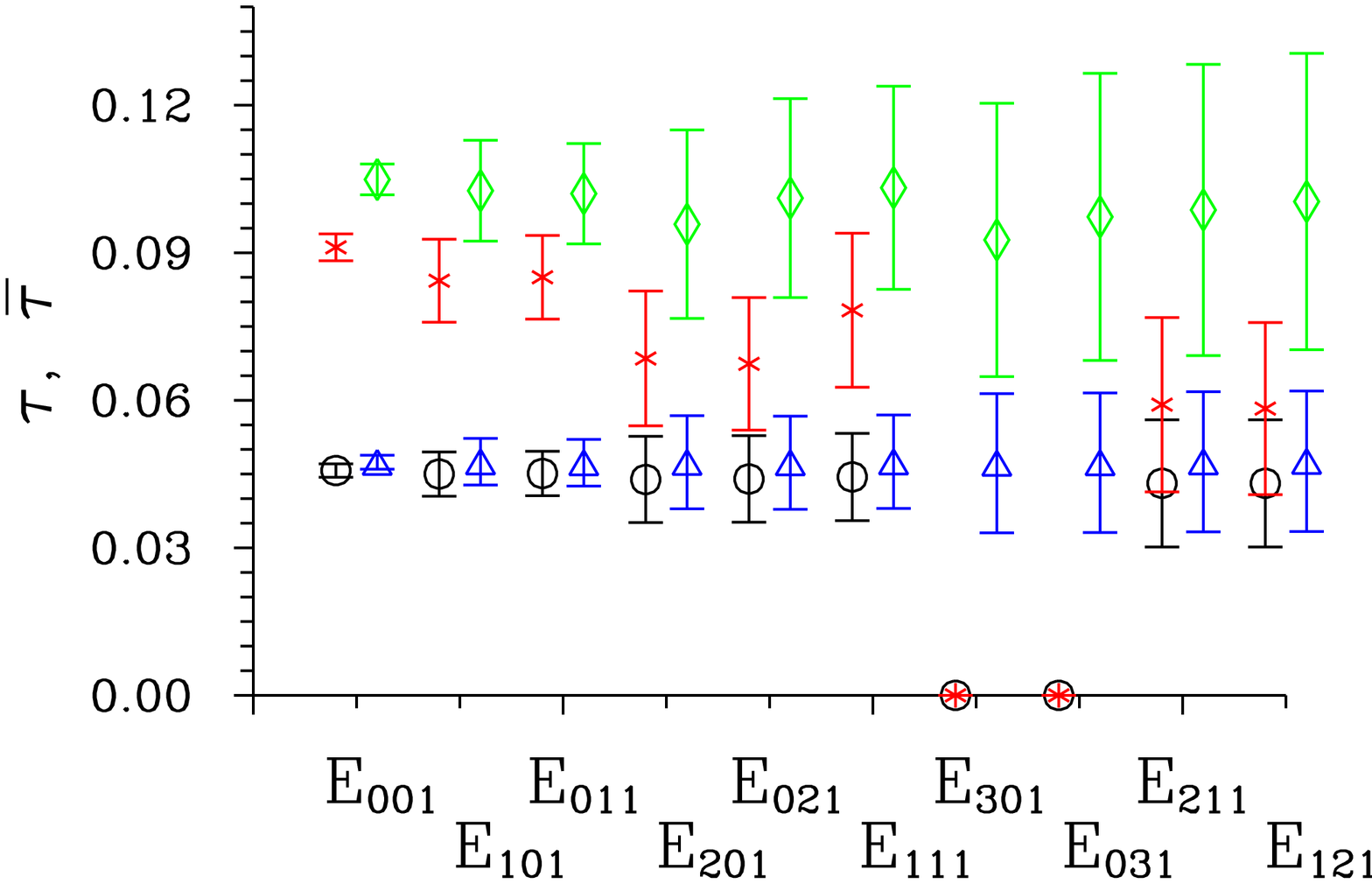}} \\
 \vspace{5mm}
 {\small (b)} \hspace{2mm}\resizebox{.8\hsize}{!}{\includegraphics{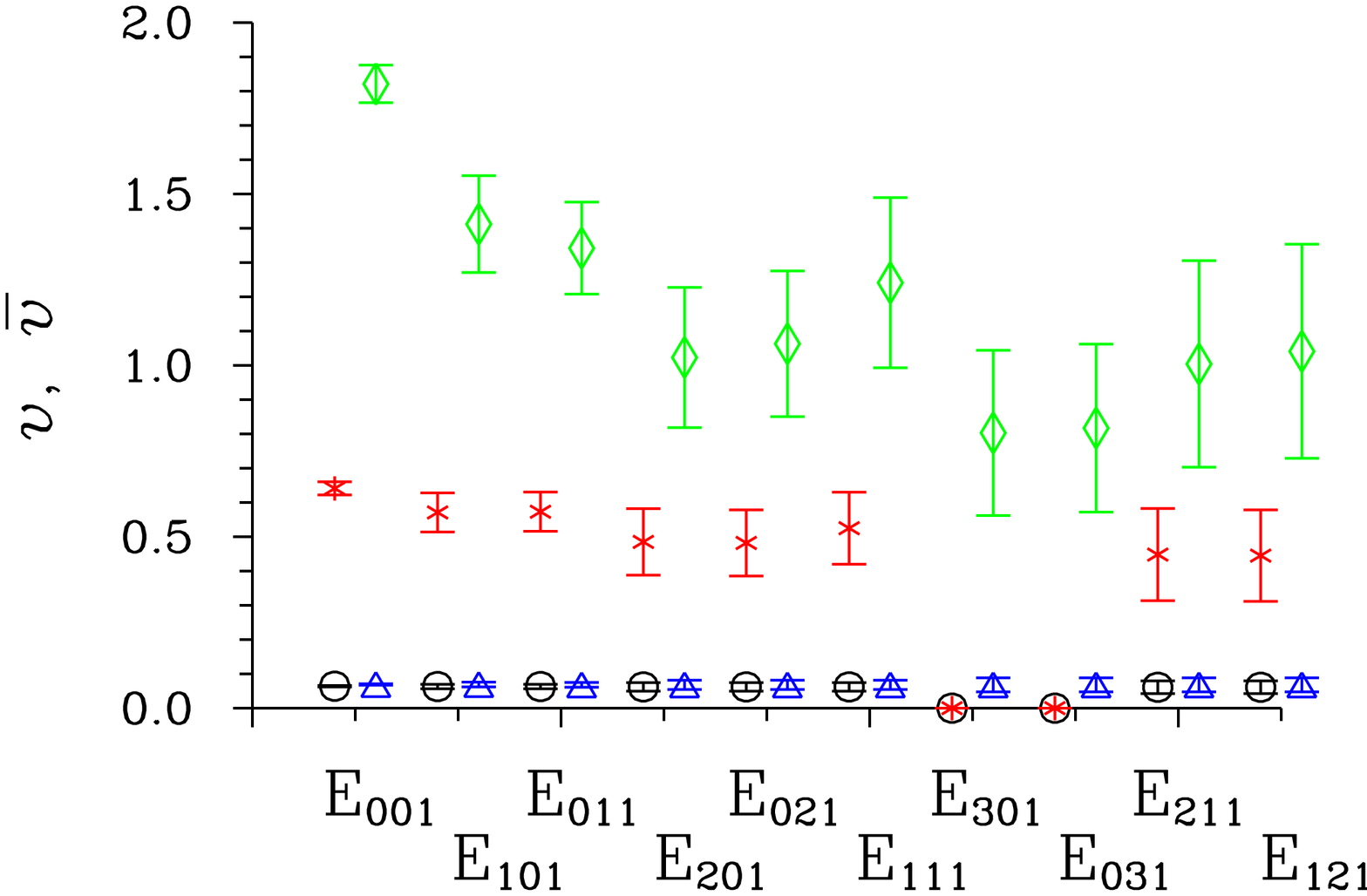}} \\
 \vspace{2mm}
 \caption{(a) Non-classicality depths $ \tau $, $ \bar{\tau} $ and (b) non-classicality
  counting parameters $ \nu $, $ \bar{\nu} $
  for NCa in Eqs.~(\ref{6}), (\ref{A1})--(\ref{A10}) based on nonnegative polynomials of
  intensities considering NCa for intensity moments ($ \tau $, $ \nu $; $ M_c=1 $: red $ \ast $, $ M_c=100 $: black $ \circ $) and
  probabilities ($ \bar{\tau} $, $ \bar{\nu} $; $ M_c=1 $: green $ \diamond $, $ M_c=100 $: blue $ \triangle $)
  and analyzing the photocount histogram.}
 \label{fig2}
\end{figure}

Next, we extend the comparison of the NCD and NCCP to a greater
(representative) set of the NCa to study the relation between the NCa for
probabilities and intensity moments. In Fig.~\ref{fig2}, we compare the
performance of NCDs and NCCPs for the NCa for probabilities written in
Eqs.~(\ref{A1})---(\ref{A10}) and corresponding to the NCa for intensity
moments (\ref{6}) up to the fifth order. We note that these NCa were suggested
in~\cite{PerinaJr2017a} for identifying and quantifying the non-classicality of
twin beams. In Fig.~\ref{fig2}, we make the comparison for two typical numbers
$ M_c $ of modes: $ M_c=1 $ and $ M_c = 100 $ [$ M_c = (M_{c,{\rm s}} +
M_{c,{\rm i}})/2 $ where $ M_{c,{\rm s}} $ and $ M_{c,{\rm i}} $ are the
numbers of modes in the signal and idler fields, respectively, determined from
the formula for multi-mode chaotic fields].
We can say in general that the NCa for probabilities give similar or better
values of the NCDs $ \bar{\tau} $ and NCCPs $ \bar{\nu} $ compared to their
counterparts in intensity moments. The NCa $ \bar{E}_{301} $ and $
\bar{E}_{031} $ for probabilities even indicate the non-classicality contrary
to the NCa $ E_{301} $ and $ E_{031} $ for intensity moments. Though the values
of NCCPs $\nu $ determined for $ M_c=1 $ are much higher than those for which
the actual number $ M_c=100 $ of field modes is applied, we suggest to use the
latter one. The reason is that the consideration of $ M_c=1 $ for multi-mode
fields may result in overestimation of the 'level of non-classicality' in the
analyzed field. For example, values of the NCD $ \tau $ greater than 1/2 may be
reached for multi-mode Gaussian fields in this case.

\begin{figure} % figs. 3a, b, c
 \resizebox{.43\hsize}{!}{\includegraphics{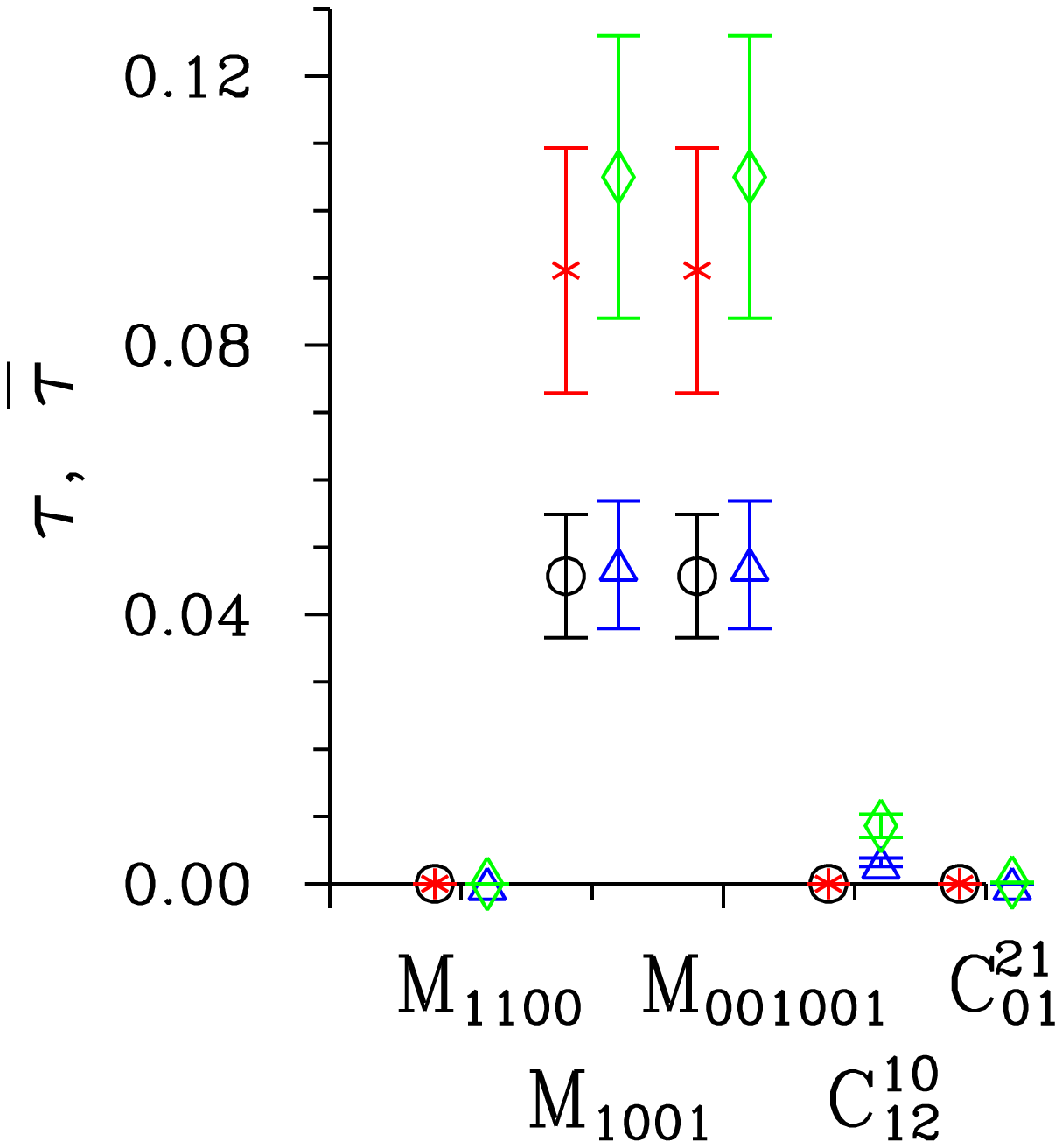}}  \hspace{1mm}
 \resizebox{.52\hsize}{!}{\includegraphics{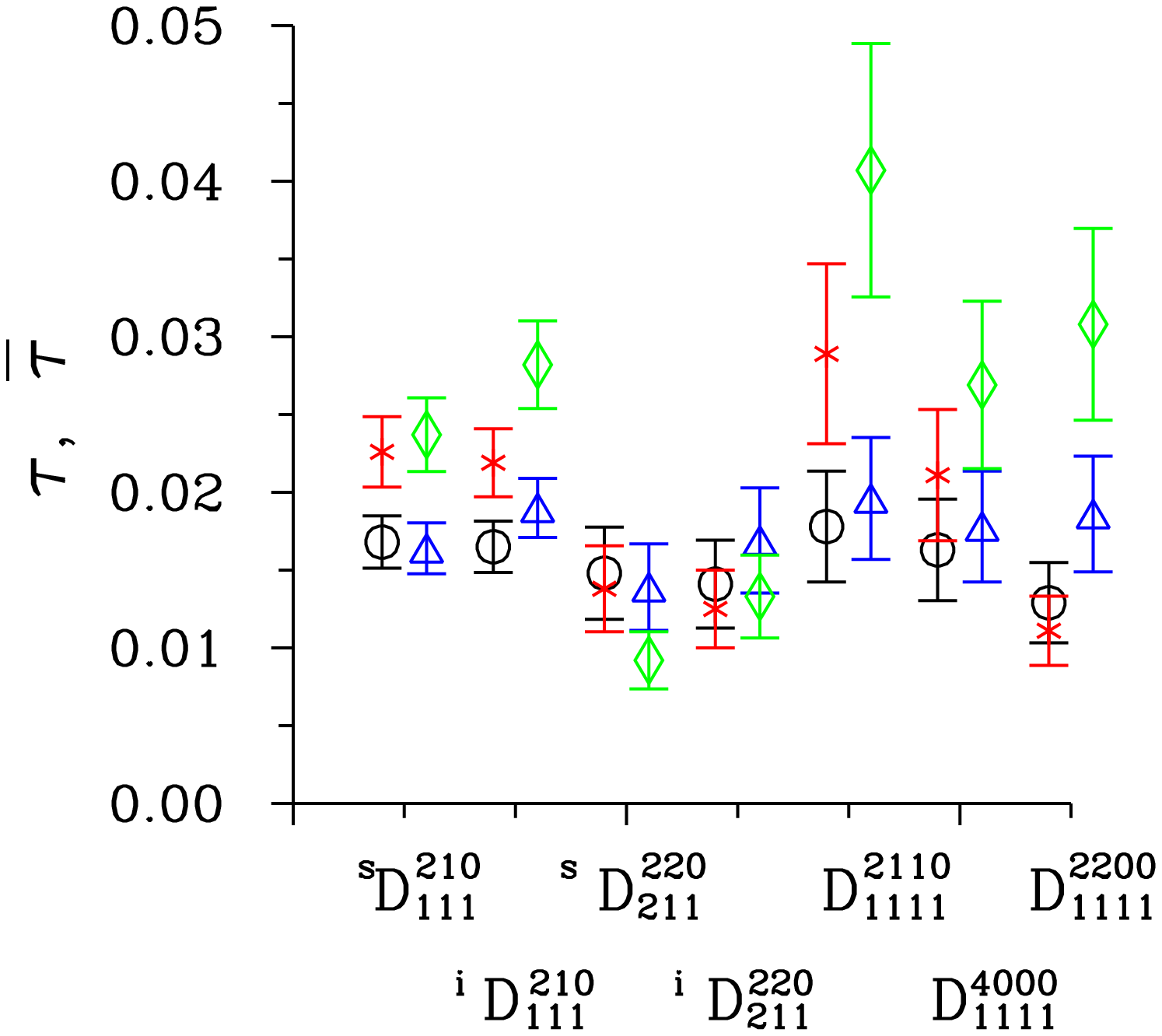}} \\
 \centerline{\small (a) \hspace{.5\hsize} (b)}
 \vspace{4mm}
 {\small (c)}\hspace{2mm}\resizebox{.8\hsize}{!}{\includegraphics{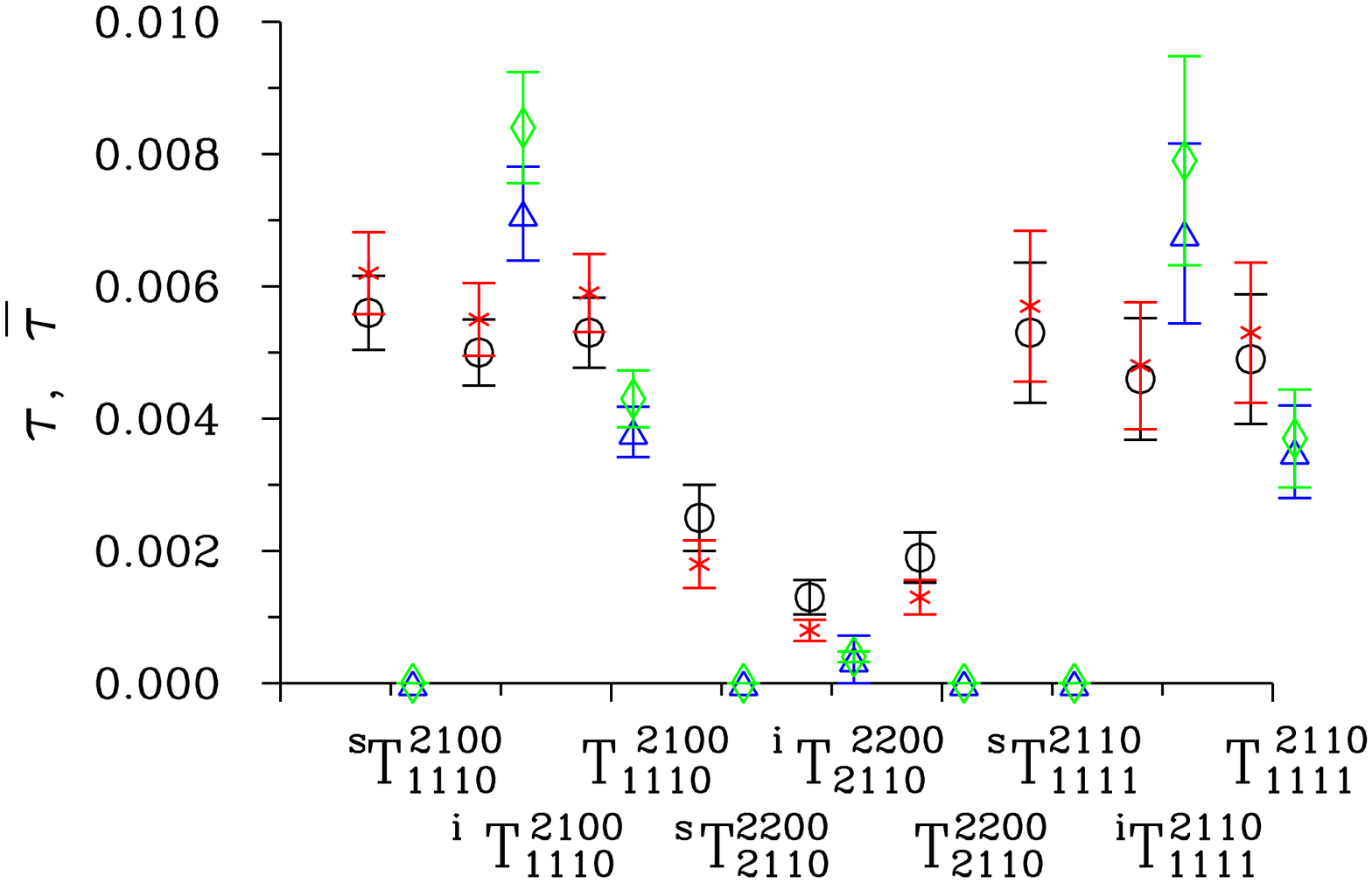}} \\
 \caption{Non-classicality depths $ \tau $, $ \bar{\tau} $ for (a) NCa in
  Eqs.~(\ref{11}), (\ref{13}), (\ref{15}), (\ref{A17})---(\ref{A21})
  derived from the matrix approach and the Cauchy--Schwarz inequality, (b) NCa in
  Eqs.~(\ref{22}), (\ref{27}), (\ref{A22})---(\ref{A26}) from the majorization theory and (c) NCa in
  Eqs.~(\ref{A27})---(\ref{A32}) from the majorization theory considering NCa for
  intensity moments ($ \tau $; $ M_c=1 $: red $ \ast $, $ M_c=100 $: black $ \circ $) and
  probabilities ($ \bar{\tau} $; $ M_c=1 $: green $ \diamond $, $ M_c=100 $: blue $ \triangle $)
  and using the photocount histogram.}
 \label{fig3}
\end{figure}

It was shown in Ref.~\cite{PerinaJr2017a} that, as a rule of thumb, the
performance of the NCa for intensity moments depends on the structure of the
terms in the NCa --- the greater the number of mean values in the product the
smaller the value of the NCD $ \tau $. The comparison of the corresponding NCa
for probabilities and intensity moments given in Fig.~\ref{fig2} suggests that
the corresponding NCa have comparable performance in revealing the
non-classicality. The mapping between the NCa for intensity moments and the NCa
for probabilities then suggests the following statement: The greater the number
of probabilities in the product, the smaller the value of the NCD $ \bar{\tau}
$. This statement is verified when we mutually compare the values of $
\bar{\tau} $ in the graphs of Fig.~\ref{fig2}(a), \ref{fig3}(a), \ref{fig3}(b),
and \ref{fig3}(c) that give the NCDs $ \bar{\tau} $ for the remaining NCa for
probabilities summarized in Appendix~A.
In more detail, the comparison of the corresponding NCa for intensity moments
and probabilities in Fig.~\ref{fig3} shows that they exhibit comparable
performance in identifying and quantifying the non-classicality, though they
operate in certain sense in complementary spaces. The NCa originating in the
majorization theory and containing terms with two intensity moments in the
product lead to systematically smaller values of the NCDs $ \tau $, as shown in
Fig.~\ref{fig3}(b). The values of NCDs $ \tau $ obtained for the NCa from the
majorization theory and having three intensity moments in the product are
systematically even smaller, as documented in Fig.~\ref{fig3}(c). In
Fig.~\ref{fig3}(c), we can see the opposite case of that discussed in relation
to the NCa $ \bar{E}_{301} $ and $ \bar{E}_{031} $: The NCa for probabilities $
^{\rm s}\bar{T}^{2100}_{1110} $, $ ^{\rm s}\bar{T}^{2200}_{2110} $, $
T^{2200}_{2110} $, and $ ^{\rm s}\bar{T}^{2110}_{1111} $ written in
Eqs.~(\ref{A27}) and (\ref{A29})---(\ref{A31}) do not indicate the
non-classicality contrary to their intensity counterparts. The discussed rule
of thumb has one exception: The NCa $ \bar{M}_{1001} $ and $ \bar{M}_{001001} $
containing two and three probabilities in the product and originating in the
matrix approach give us the values of NCDs $ \bar{\tau} $ comparable with the
greatest values of $ \bar{\tau} $ reached by the NCa derived from nonnegative
polynomials of intensities and plotted in Fig.~\ref{fig2}(a).

We note that the NCa $ \bar{E}_{002} $, $ \bar{E}_{102} $, $ \bar{E}_{012} $, $
\bar{E}_{1011} $, $ \bar{E}_{0111} $ and $ \bar{E}_{0011} $ defined for
probabilities in Eqs.~(\ref{A11})---(\ref{A16}) do not reveal the
non-classicality of the analyzed twin beam, similarly as their intensity
counterparts \cite{PerinaJr2017a}.

\section{Application of non-classicality criteria for arbitrarily large photon numbers}

In this section, we demonstrate the application of the parametric systems of
the NCa for probabilities derived in Sec.~II to the experimental twin beam
characterized by the photocount histogram $ f_{\rm si}(c_{\rm s},c_{\rm i}) $
used in Sec.~IV. We apply the derived parametric systems of the NCa in three
scenarios to investigate the non-classicality. First, we only consider the
values of the indices in the NCa such that the corresponding pairs (signal and
idler) of photon numbers lie in specific well-defined areas. For example,
strips parallel to the diagonal $ n_{\rm s} = n_{\rm i} $ are suitable for the
non-classicality analysis of twin beams with their photon-pair-like structure.
Second, we impose specific conditions for the values of indices. Consideration
of the NCa that include only probabilities of the neighbor photon numbers may
serve as an example. The NCa originating in the majorization inequalities in
which 'one ball is moved' also belong to this category. Third, we may consider
a whole parametric system of the NCa, scan over all allowed values of the
indices and look for the maximal value of the chosen non-classicality
quantifier. The first two scenarios may provide us the information about
'location of the non-classicality' across the photon-number distribution. This
property allows us to get also certain information about the quality of the
analyzed beam, especially about the present noise and its distribution across
the photon-number distribution. To demonstrate this property, we apply the
parametric systems of the NCa in parallel to three photon-number distributions.
We consider two reconstructed photon-number distributions $ p_{\rm si}(n_{\rm
s},n_{\rm i}) $ of the analyzed twin beam: The first one was obtained by the
maximum likelihood approach~\cite{Dempster1977,PerinaJr2013a}. The second
reconstructed photon-number distribution is a best fit of the experimental
photocount histogram by a multi-mode Gaussian twin beam with three multi-mode
components describing ideally paired photons and noisy signal and noisy idler
photons \cite{PerinaJr2012a,PerinaJr2013a,Perina2005}. The third considered
photon-number distribution characterizes an ideal (noiseless) twin beam with
the mean photon-pair number equal to that of the reconstructed twin beam (8.86
mean photon pairs equally divided into 80 modes). Details about the
reconstruction are given in Ref.~\cite{PerinaJr2017a} where the reconstructed
photon-number distributions are plotted in Figs.~2(b,c).

\begin{figure} % figs. 4a, b, c, d, e, f
 \resizebox{.47\hsize}{!}{\includegraphics{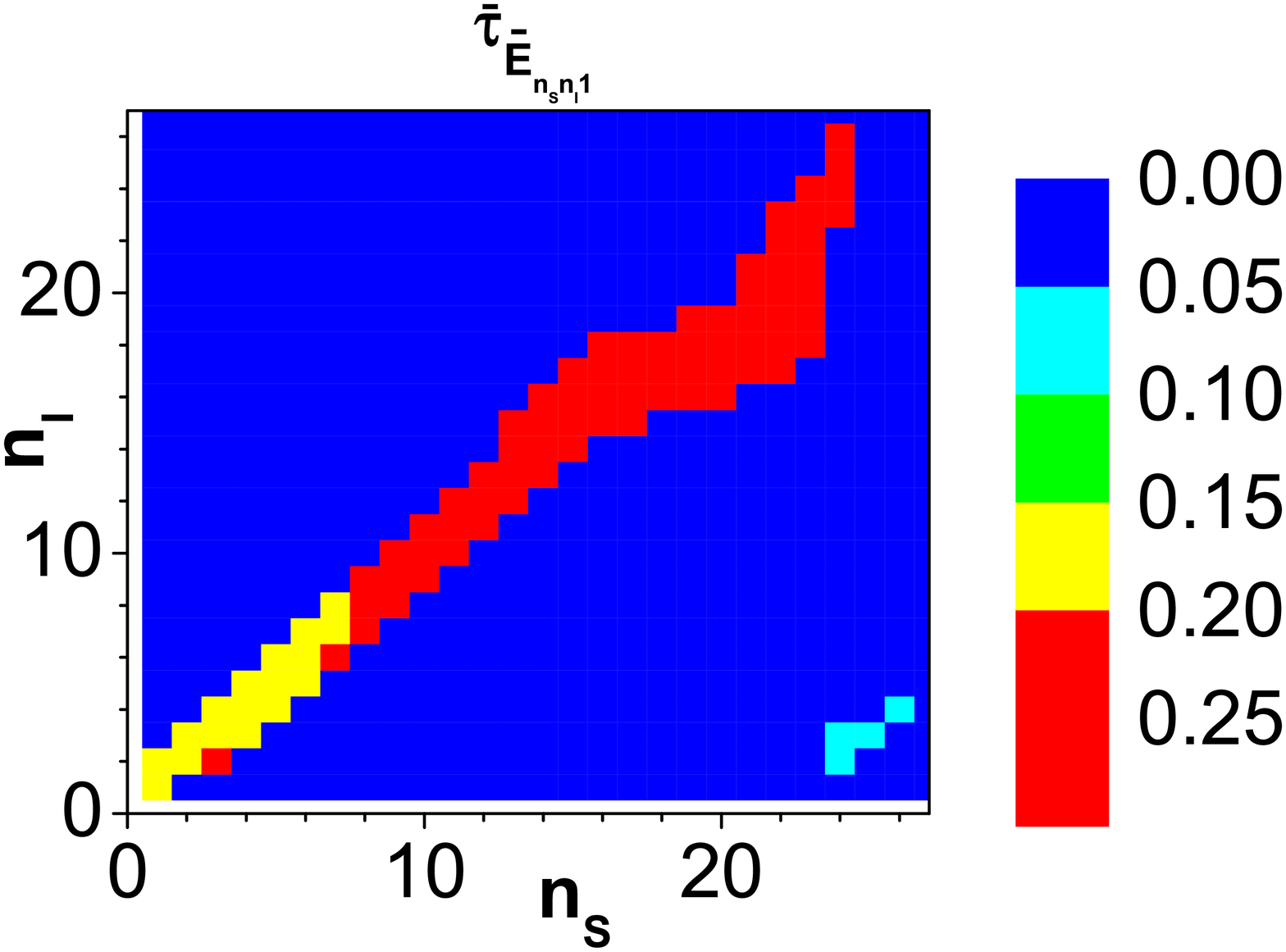}} \hspace{2mm}
 \resizebox{.47\hsize}{!}{\includegraphics{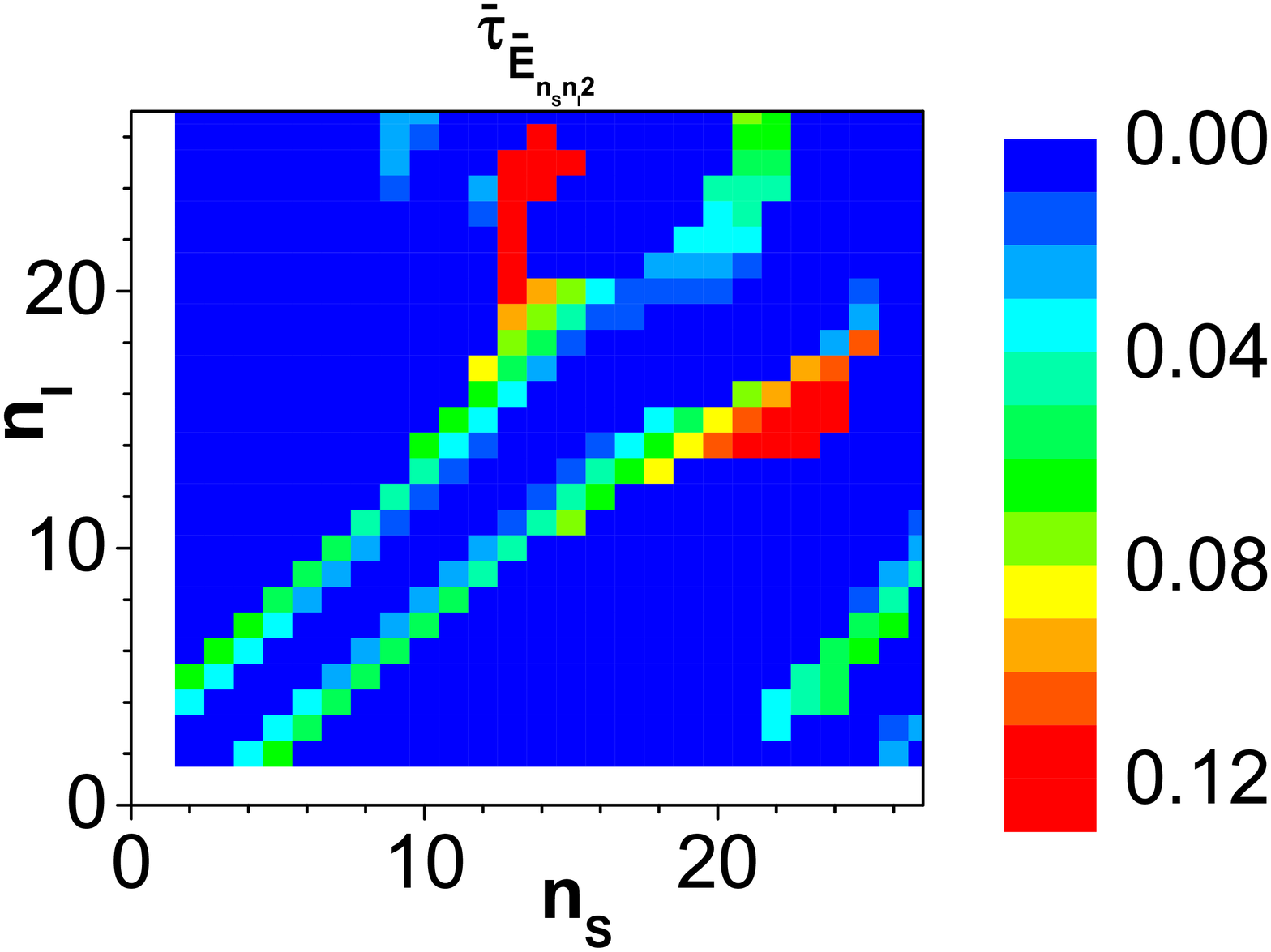}} \\
 \centerline{\small (a) \hspace{.4\hsize} (b)}
  \vspace{1mm}
 \resizebox{.47\hsize}{!}{\includegraphics{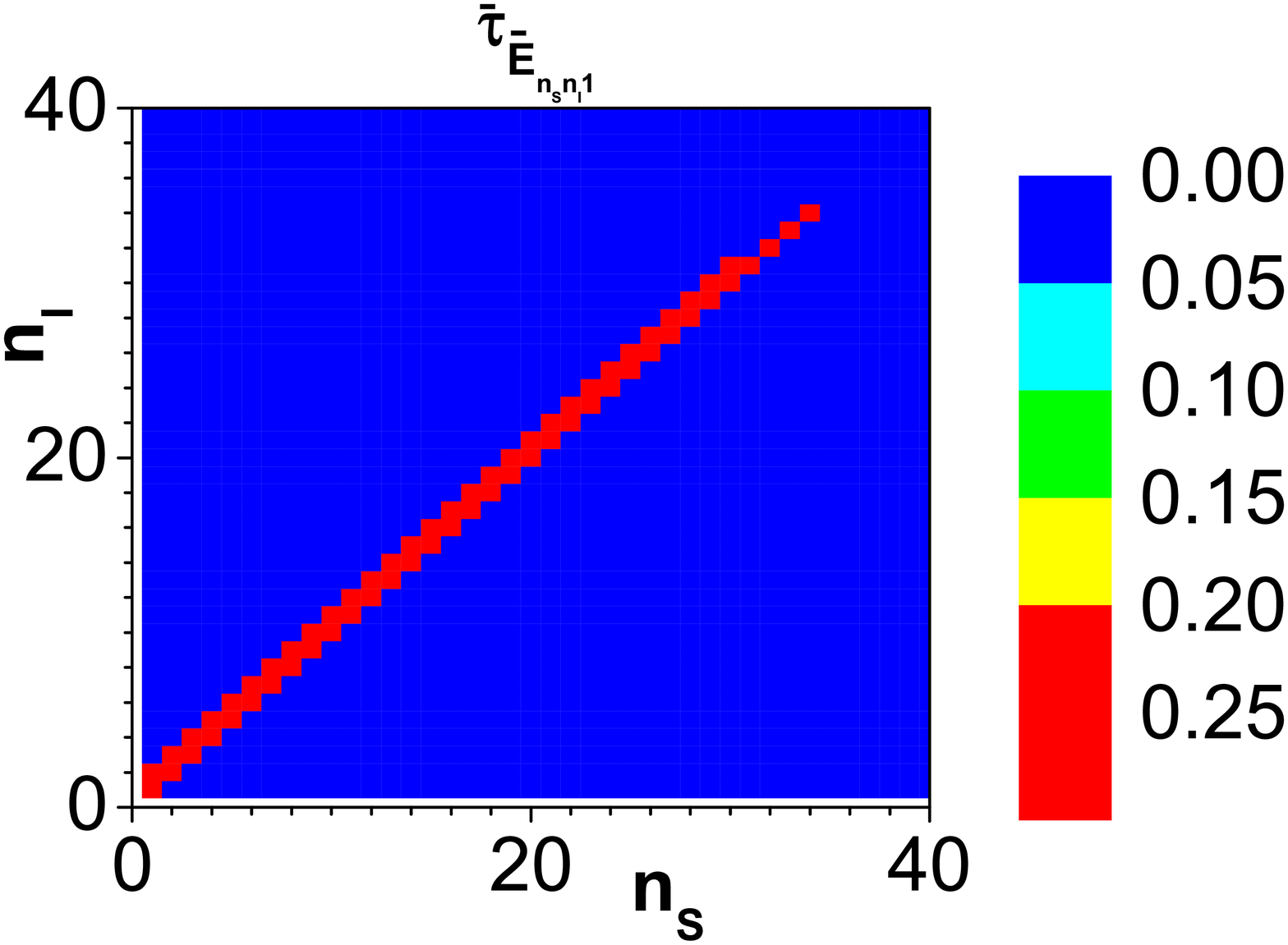}} \hspace{2mm}
 \resizebox{.47\hsize}{!}{\includegraphics{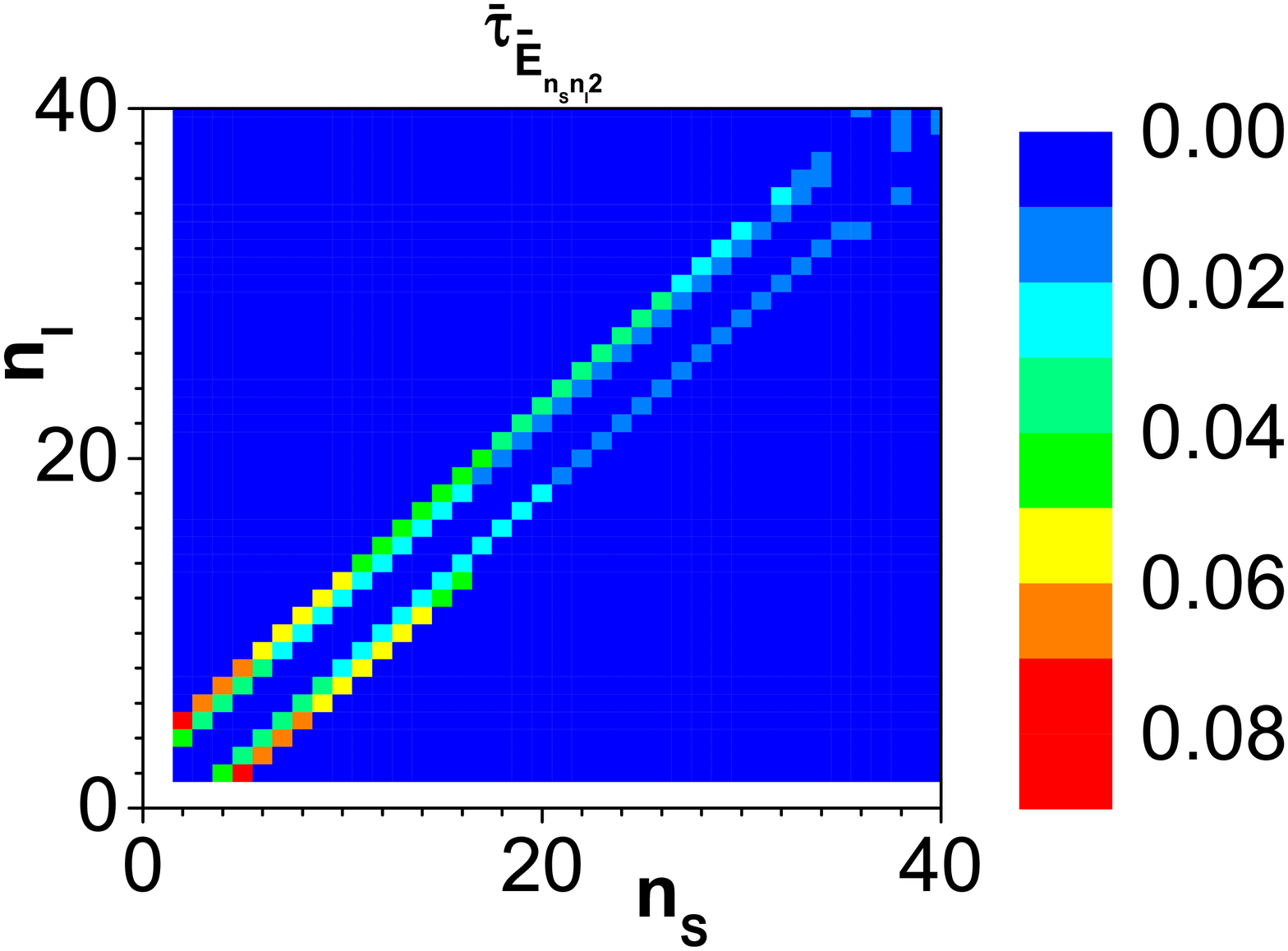}} \\
 \centerline{\small (c) \hspace{.4\hsize} (d)}
  \vspace{1mm}
 \resizebox{.47\hsize}{!}{\includegraphics{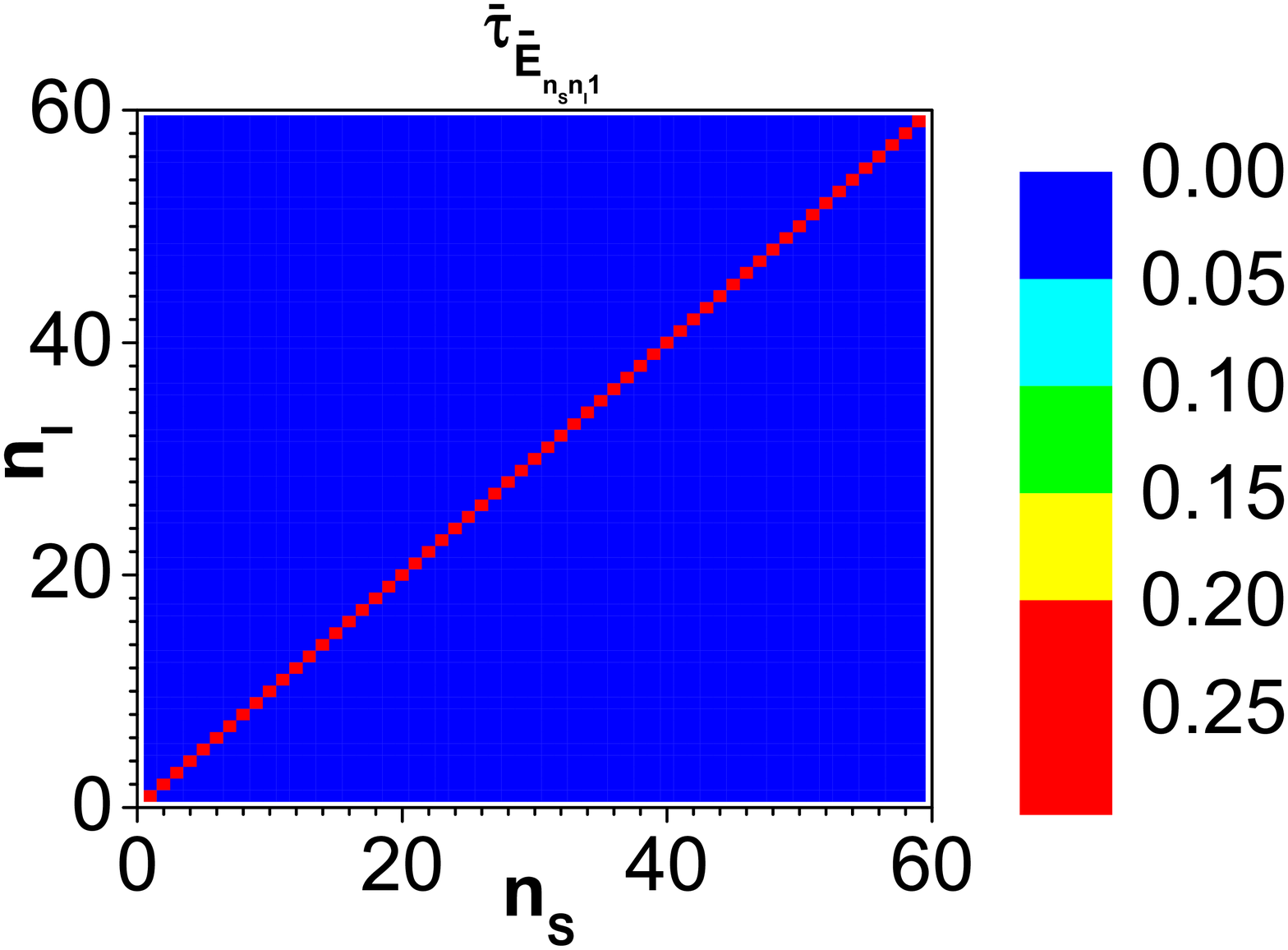}} \hspace{2mm}
 \resizebox{.47\hsize}{!}{\includegraphics{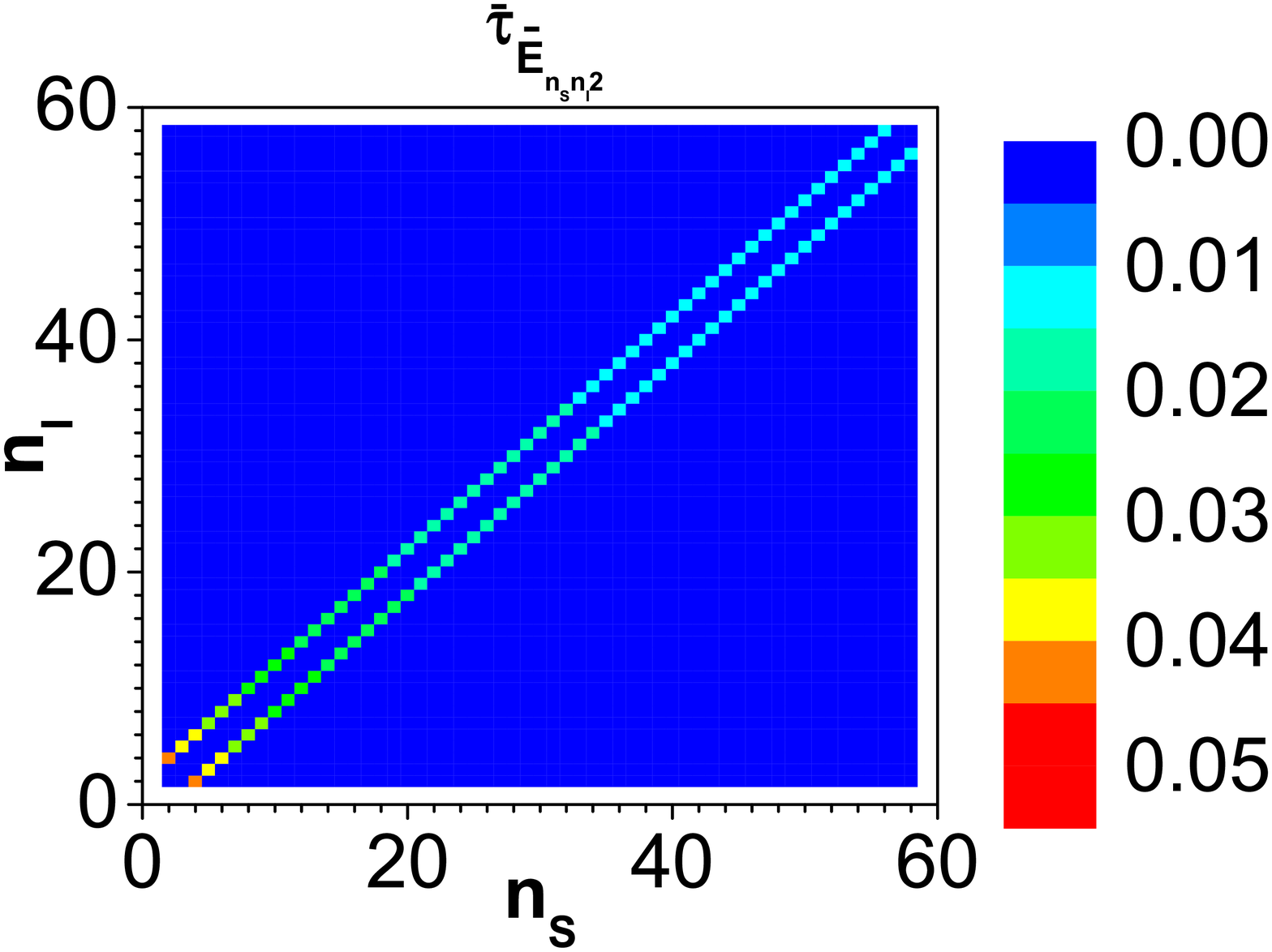}} \\
 \centerline{\small (e) \hspace{.4\hsize} (f)}
 \caption{Non-classicality depths $ \bar{\tau} $ of NCa $ \bar{E}_{n_{\rm s}n_{\rm i}1} $ in Eq.~(\ref{6}) (a,c,e) and
  $ \bar{\tau} $ of NCa $ \bar{E}_{n_{\rm s}n_{\rm i}2} $ (b,d,f) in Eq.~(\ref{7}) as they depend on indices $ n_{\rm s} $
  and $ n_{\rm i} $
  for photon-number distributions obtained by reconstruction using the maximum-likelihood approach (a,b)
  and Gaussian fit (c,d) and characterizing an ideal twin beam with 8.86 mean photon pairs equally divided into 80 modes
  (e,f); $ M_n = 80 $. Relative experimental errors are lower than 3~\% in (a,c) and 5~\% (b,d).}
 \label{fig4}
\end{figure}

The NCa $ \bar{E}_{n_{\rm s}n_{\rm i}1} $ and $ \bar{E}_{n_{\rm s}n_{\rm i}2} $
defined in Eqs.~(\ref{6}) and (\ref{7}) may serve in the first scenario. They
allow to reveal the photon-pair-like structure of twin beams, as documented by
the corresponding NCDs $ \bar{\tau} $ plotted in Fig.~\ref{fig4} for the above
mentioned photon-number distributions. Whereas the NCa $ \bar{E}_{n_{\rm
s}n_{\rm i}1} $ identify as nonclassical the area with photon numbers $ n_{\rm
s} $ and $ n_{\rm i} $ at and close to the diagonal $ n_{\rm s}=n_{\rm i} $
[see Figs.~\ref{fig4}(a,c,e)], the NCa $ \bar{E}_{n_{\rm s}n_{\rm i}2} $ assign
considerably weaker non-classicality to two strips at the sides of the diagonal
[see Figs.~\ref{fig4}(b,d,f)].
The comparison of graphs in Figs.~\ref{fig4}(a) [(b)] and (c) [(d)] drawn for
the reconstructed photon-number distributions with that in Fig.~\ref{fig4}(e)
[(f)] for the ideal twin beam shows that the reconstruction based on the
Gaussian fit is much less noisy than the maximum-likelihood reconstruction.
Despite different amount of the noise present in three distributions in
Fig.~\ref{fig4}, the determined NCDs $ \bar{\tau} $ are comparable both for the
NCa $ \bar{E}_{n_{\rm s}n_{\rm i}1} $ and the NCa $ \bar{E}_{n_{\rm s}n_{\rm
i}2} $. The maximal NCDs $ \bar{\tau} $ reach 0.25 in case of $ \bar{E}_{n_{\rm
s}n_{\rm i}1} $, they fluctuate around 0.08 for $ \bar{E}_{n_{\rm s}n_{\rm i}2}
$. It is worth noting that according to the graphs in Figs.~\ref{fig4}(a) and
(b) the photon-number distribution from the maximum likelihood approach partly
loses its diagonal-like structure for greater photon numbers $ n_{\rm s} $ and
$ n_{\rm i} $. This probably occurs as a consequence of the experimental noise
and its amplification in the reconstruction algorithm.

\begin{figure} % figs. 5a, b
 \resizebox{.47\hsize}{!}{\includegraphics{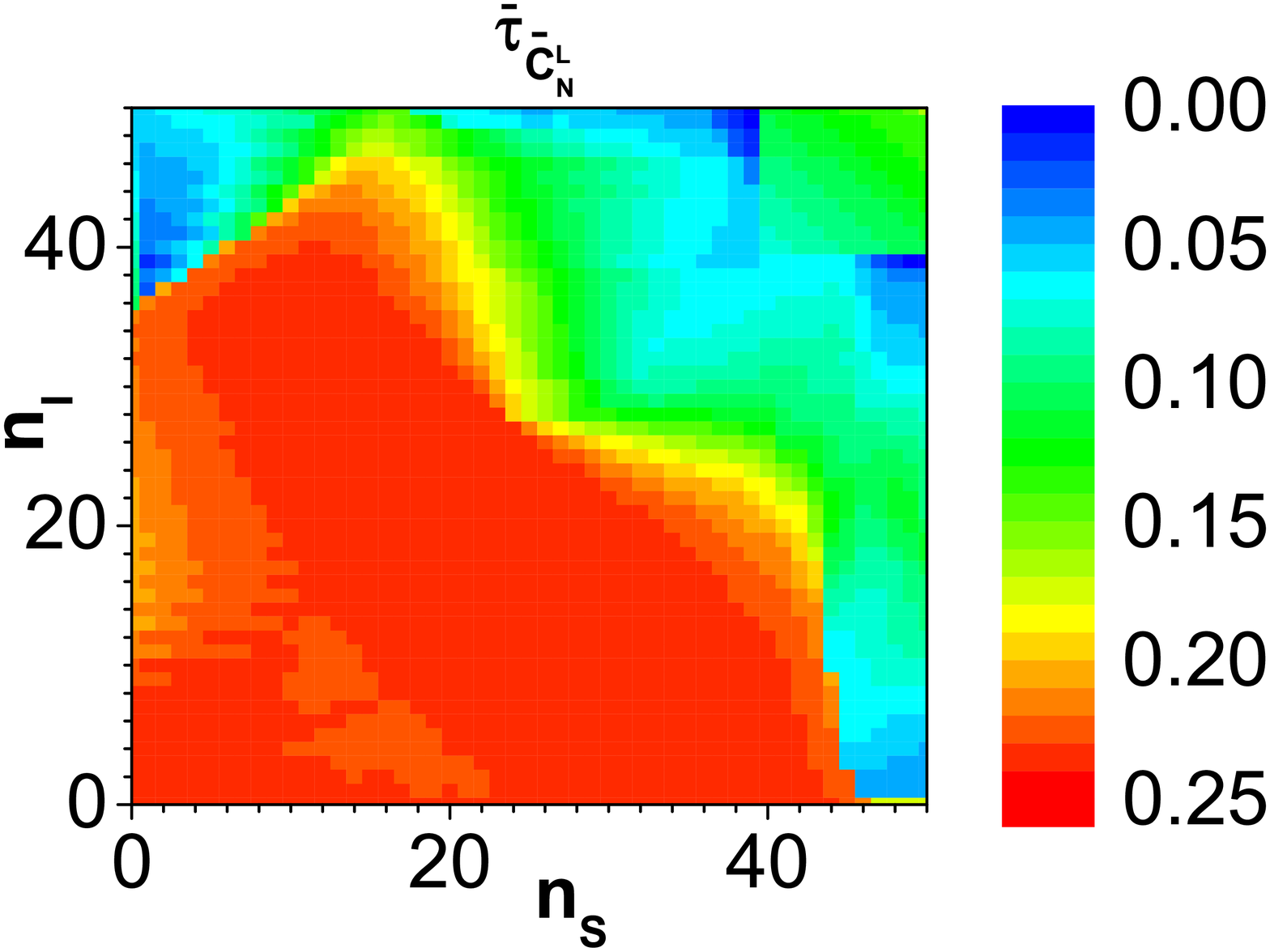}} \hspace{2mm}
 \resizebox{.47\hsize}{!}{\includegraphics{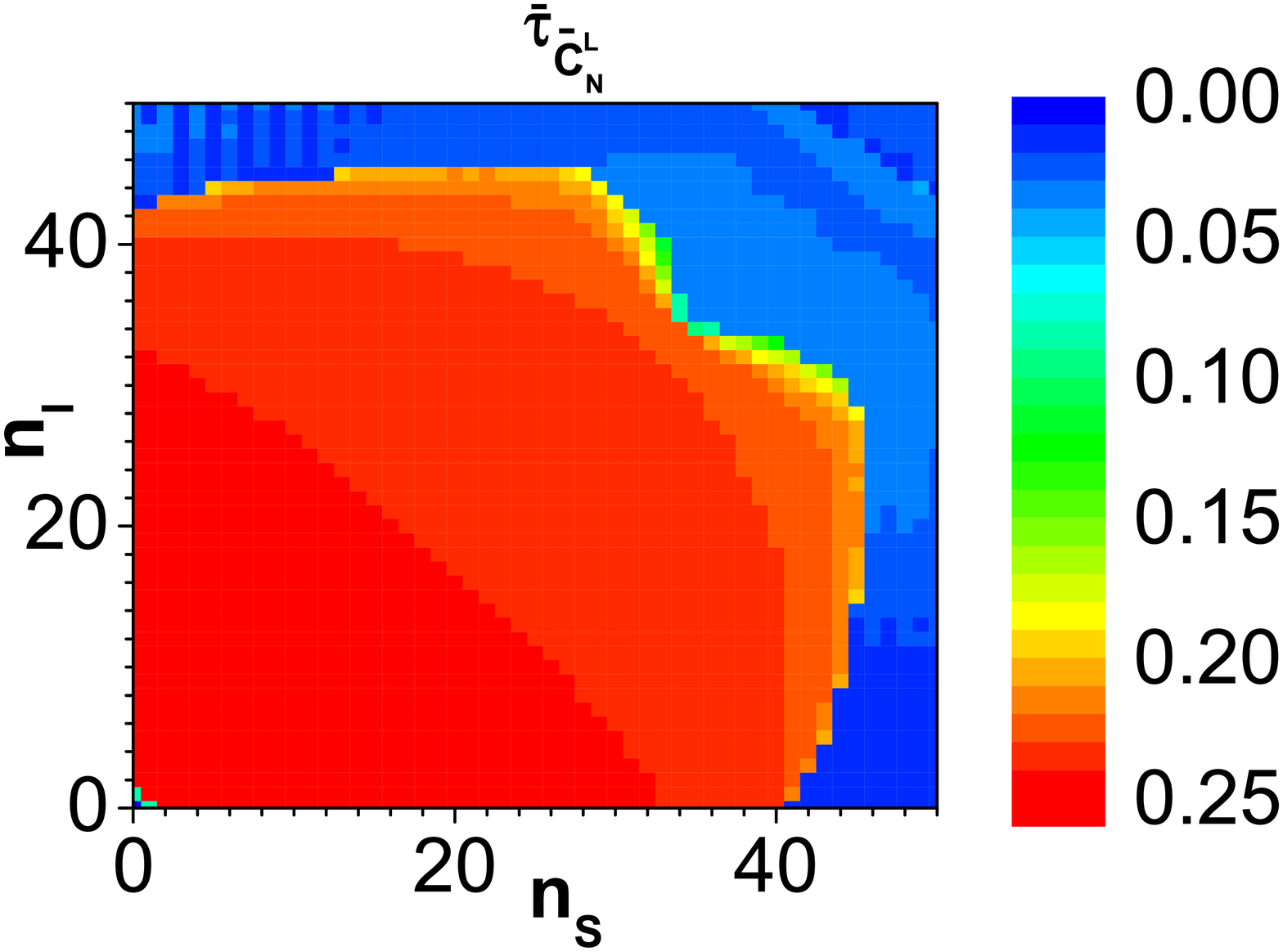}} \\
 \centerline{\small (a) \hspace{.4\hsize} (b)}
 \caption{Maximal non-classicality depth $ \bar{\tau} $ of NCa $ \bar{C}_{N}^{L} $ in Eq.~(\ref{12}) for arbitrary $ L, N $
  as a function of signal ($ n_{\rm s} $) and idler ($ n_{\rm i} $) photon numbers for photon-number distributions
  obtained by the reconstruction using (a) the maximum-likelihood approach and (b) the Gaussian fit; $ M_n = 80 $.
  Relative experimental errors are lower than 6~\%.}
 \label{fig5}
\end{figure}

The NCa $ \bar{C}_{N}^{L} $ in Eq.~(\ref{12}) derived from the Cauchy-Schwarz
inequality allow to identify the non-classicality in a wide area of the
photon-number distributions: According to the graphs in Fig.~\ref{fig5} the
probabilities for photon-numbers $ n_{\rm s} $ and $ n_{\rm i} $ approximately
fulfilling the condition $ n_{\rm s} + n_{\rm i} \le 40 $ enter into some NCa
whose NCDs $ \bar{\tau} \approx 0.25 $. These values of $ \bar{\tau} $ are
comparable to the greatest values attained in the analysis. We can see also
here that the graph in Fig.~\ref{fig5}(b) for the Gaussian fit is much more
symmetric/regular than that in Fig.~\ref{fig5}(a) originating in the
maximum-likelihood approach. We note that greater values of $ \bar{\tau}
\approx 0.25 $ occur also for greater photon numbers $ n_{\rm s} $ and $ n_{\rm
i} $ because there is an average shift in photon numbers $ M_n\bar{\tau} $
equal to 20 for $ \bar{\tau}=0.25 $.

\begin{figure} % figs. 6a, b
 \resizebox{.47\hsize}{!}{\includegraphics{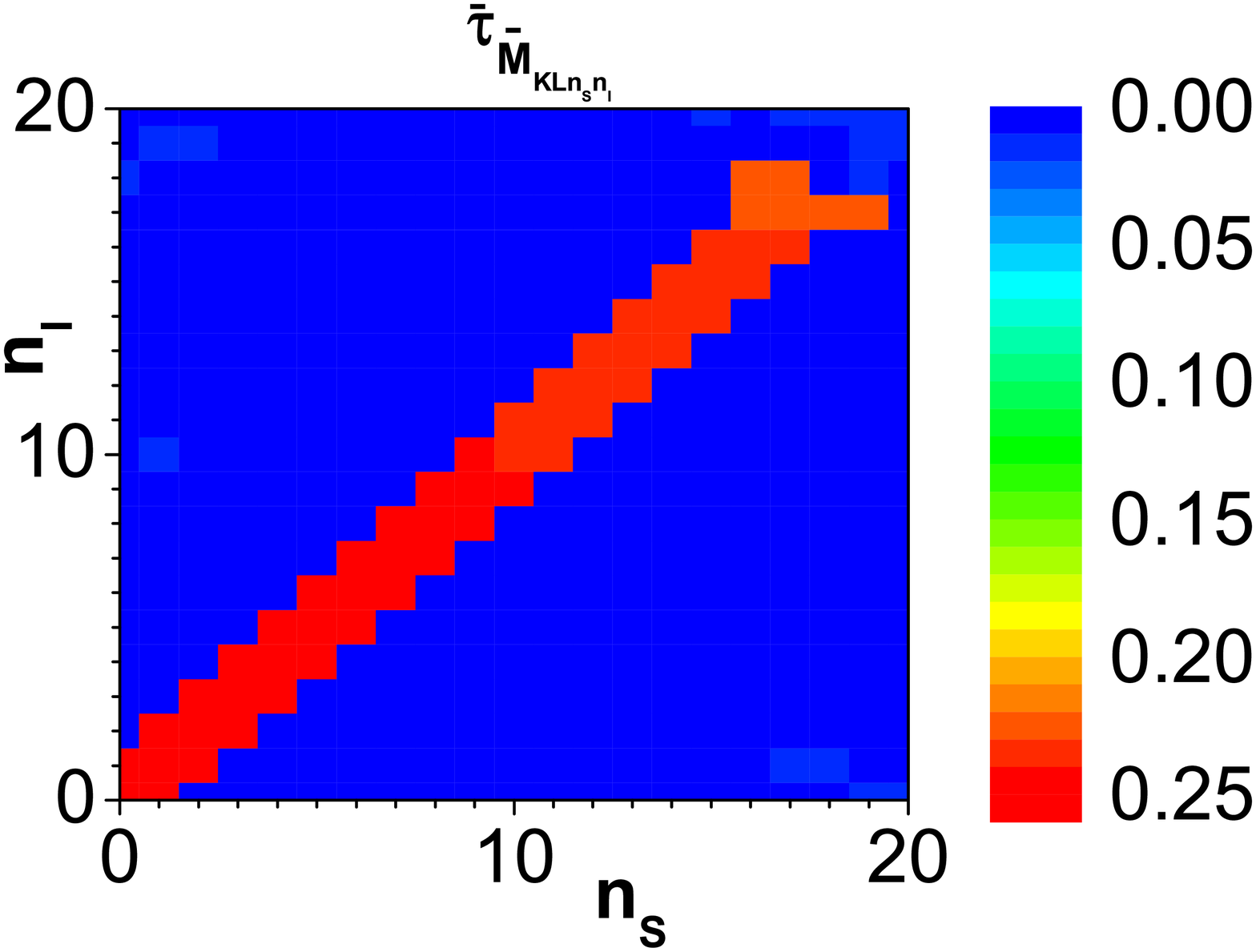}} \hspace{2mm}
 \resizebox{.47\hsize}{!}{\includegraphics{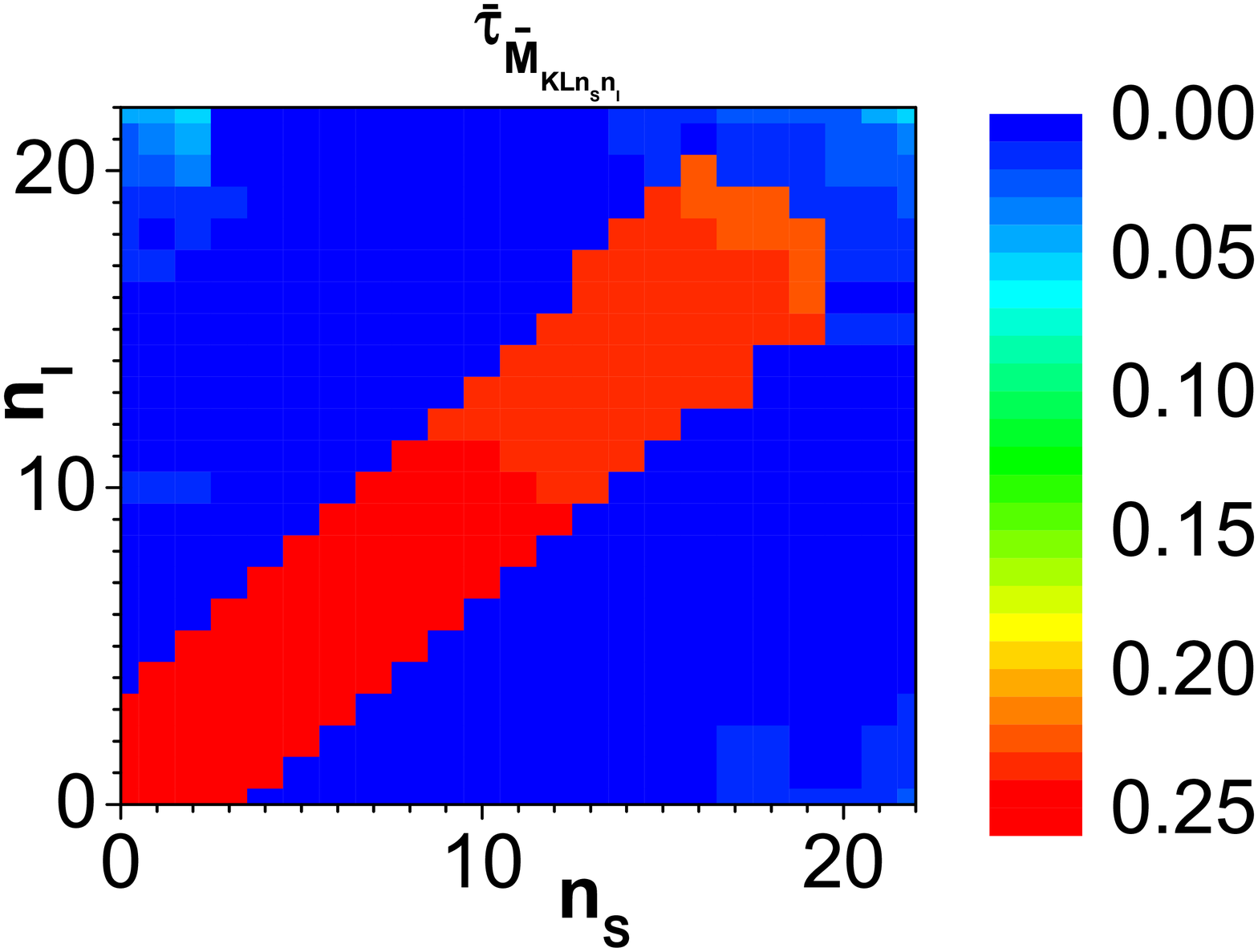}} \\
 \centerline{\small (a) \hspace{.4\hsize} (b)}
 \caption{Maximal non-classicality depth $ \bar{\tau} $ of NCa $ \bar{M}_{KLN} $ in Eq.~(\ref{16}) for (a) [(b)]
  $ K $ and $ L $
  differing from $ N $ by maximally $ d=1 $ [2] ($ |k_a - n_a| \le d $, $ |l_a - n_a| \le d $ for $ a = {\rm s,i} $)
  as it depends on signal ($ n_{\rm s} $) and idler ($ n_{\rm i} $) photon numbers for photon-number distribution
  reached by the maximum-likelihood approach; $ M_n = 80 $.
  Relative experimental errors are lower than 9~\%.}
 \label{fig6}
\end{figure}

The NCa $ \bar{M}_{KLN} $ in Eq.~(\ref{16}) with six indices can be utilized
for indicating the 'local' non-classicality. For this use, we apply the
condition for relative differences in the values of the corresponding indices.
The chosen maximal value of the difference gives us 'the resolution' in the
analysis. This may be useful for more intense fields. The graphs in
Figs.~\ref{fig6}(a) and \ref{fig6}(b) drawn for two adjacent resolutions show
that broadening of the neighborhood in which the probabilities reside may
considerably broaden the area in the plane $ (n_{\rm s}, n_{\rm i}) $ where the
non-classicality is observed. The fact that the corresponding NCDs $ \bar{\tau}
\approx 0.25 $, i.e. close to the maximal attained values, reveals the NCa $
\bar{M}_{KLN} $ as suitable for quantifying the 'local' non-classicality.

The parametric systems of the NCa derived from the majorization theory are
applicable in the second scenario. The imposed condition is given by the number
$ m $ of 'balls' moved when constructing given NCa. This number quantifies the
relative change in the values of the involved indices. These indices then label
the used signal and idler photon numbers $ n_{\rm s} $ and $ n_{\rm i} $. In
the below examples involving the NCa $ ^{\rm s}\bar{\cal D}_{kl}^{2,m} $ in
Eq.~(\ref{25}) and $ \bar{\cal D}_{kl}^{3,m} $ in Eq.~(\ref{31}), we show that
these parametric systems may behave complementary and they are sensitive to the
presence of the 'local' noise.

\begin{figure} % figs. 7a, b
 \resizebox{.47\hsize}{!}{\includegraphics{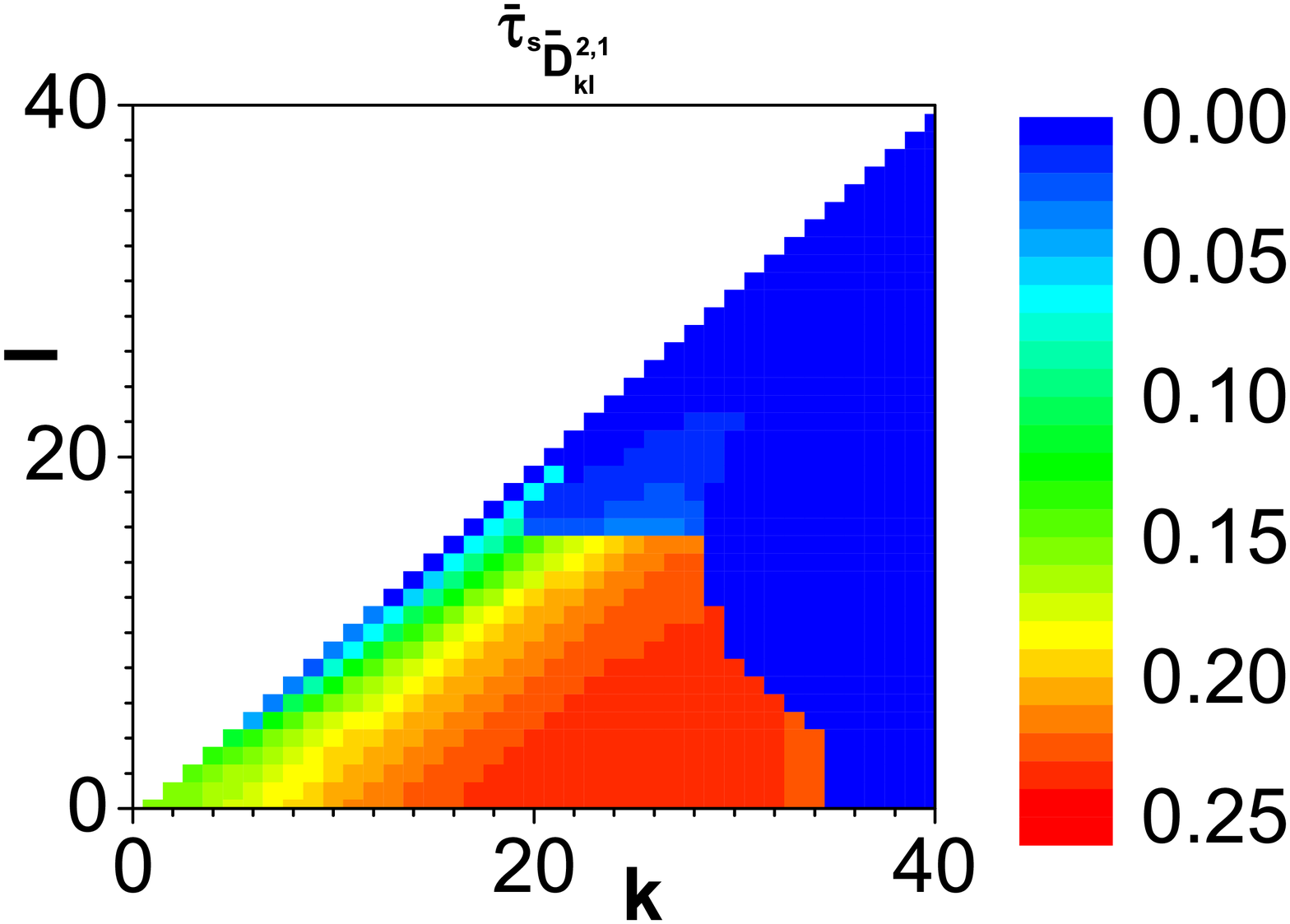}} \hspace{2mm}
 \resizebox{.47\hsize}{!}{\includegraphics{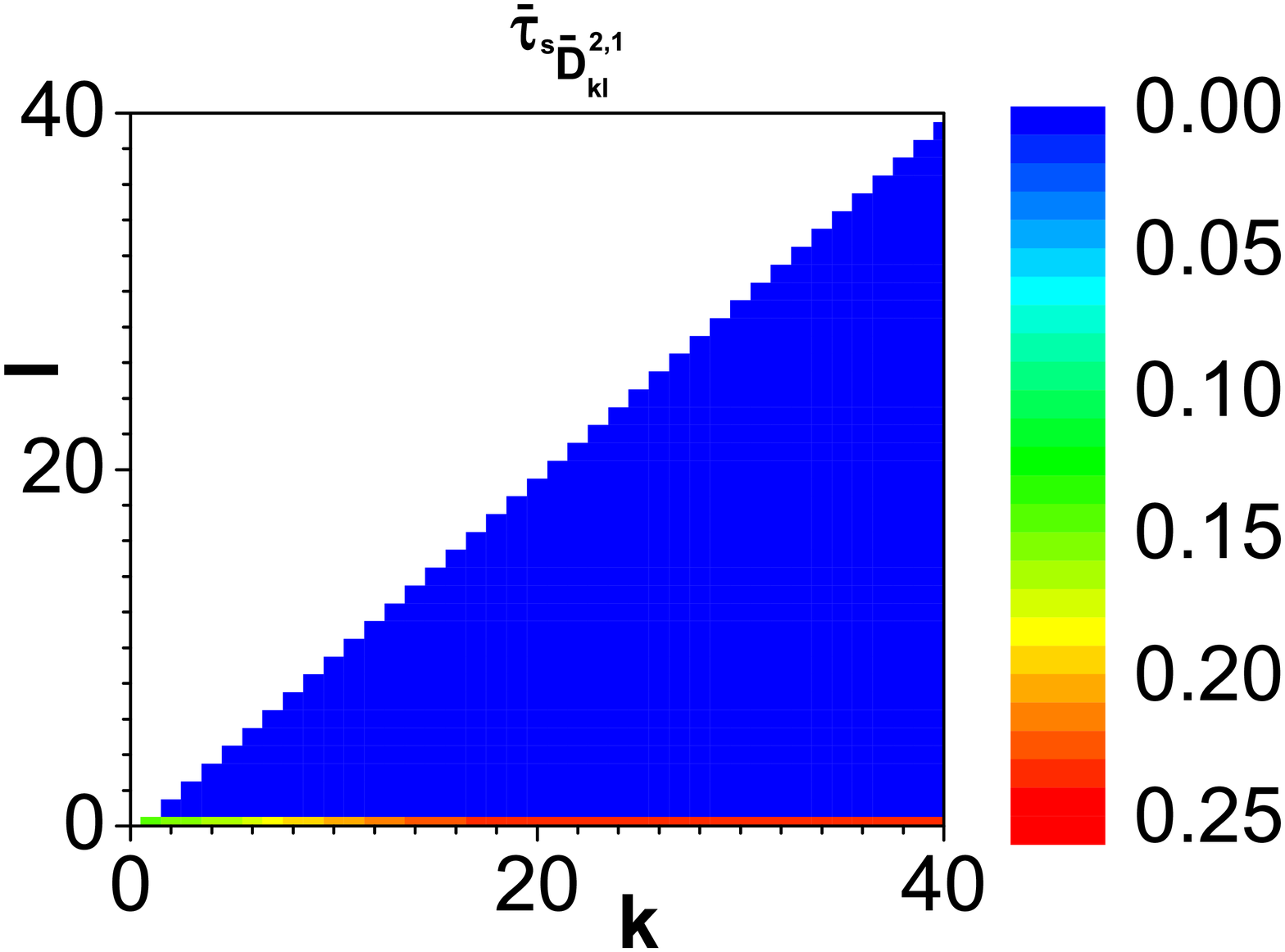}} \\
 \centerline{\small (a) \hspace{.4\hsize} (b)}
 \caption{Non-classicality depth $ \bar{\tau} $ of NCa $ ^{\rm s}\bar{\cal D}_{kl}^{2,1} $ in Eq.~(\ref{25}) as it depends on
  indices $ k $ and $ l $ for (a) photon-number distribution giving the Gaussian fit and (b) ideal twin beam with 8.86
  mean photon pairs equally divided into 80 modes; $ M_n = 80 $.
  Relative experimental errors are lower than 6~\%.}
 \label{fig7}
\end{figure}

\begin{figure} % figs. 8a, b
 \resizebox{.47\hsize}{!}{\includegraphics{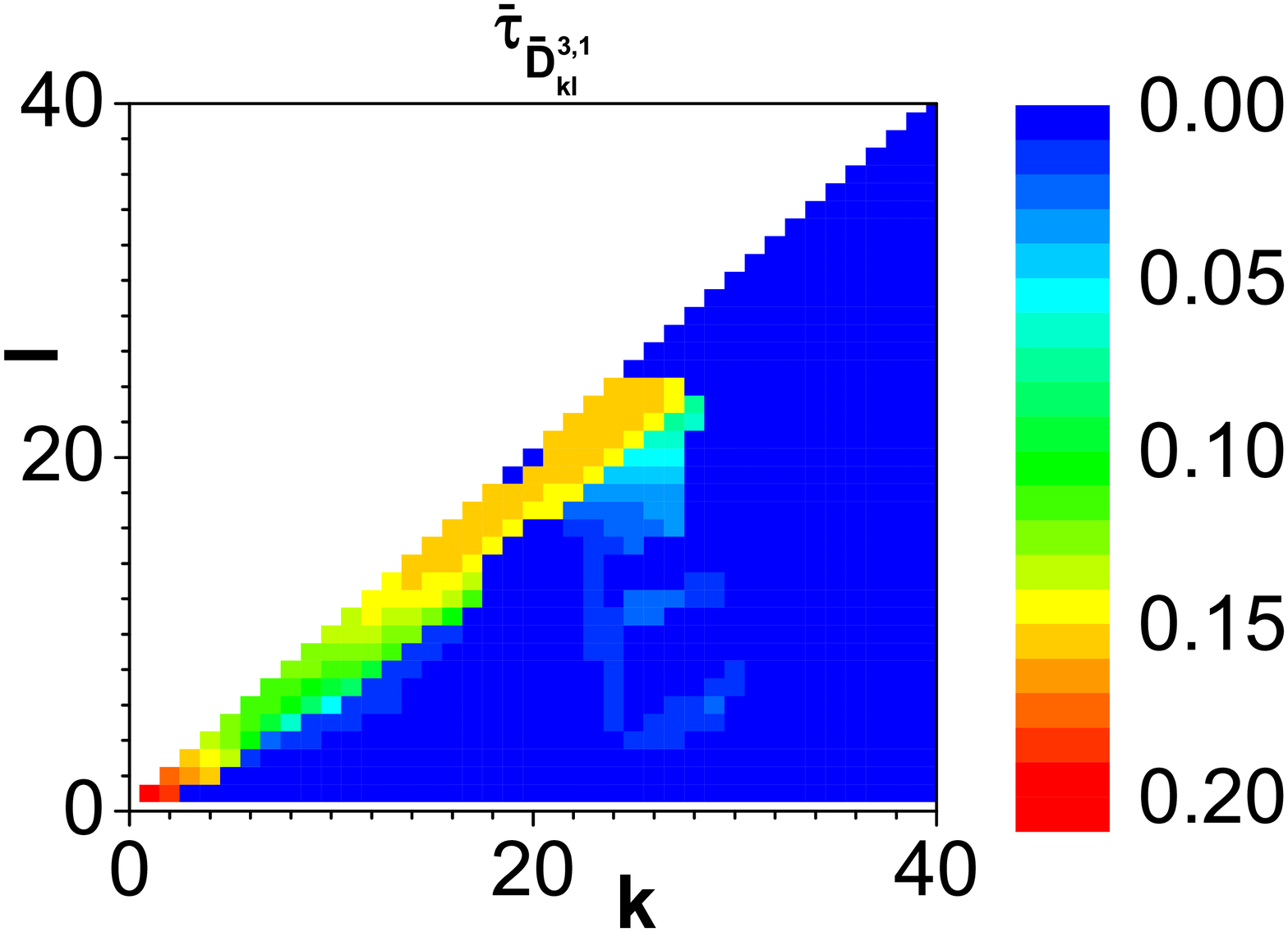}} \hspace{2mm}
 \resizebox{.47\hsize}{!}{\includegraphics{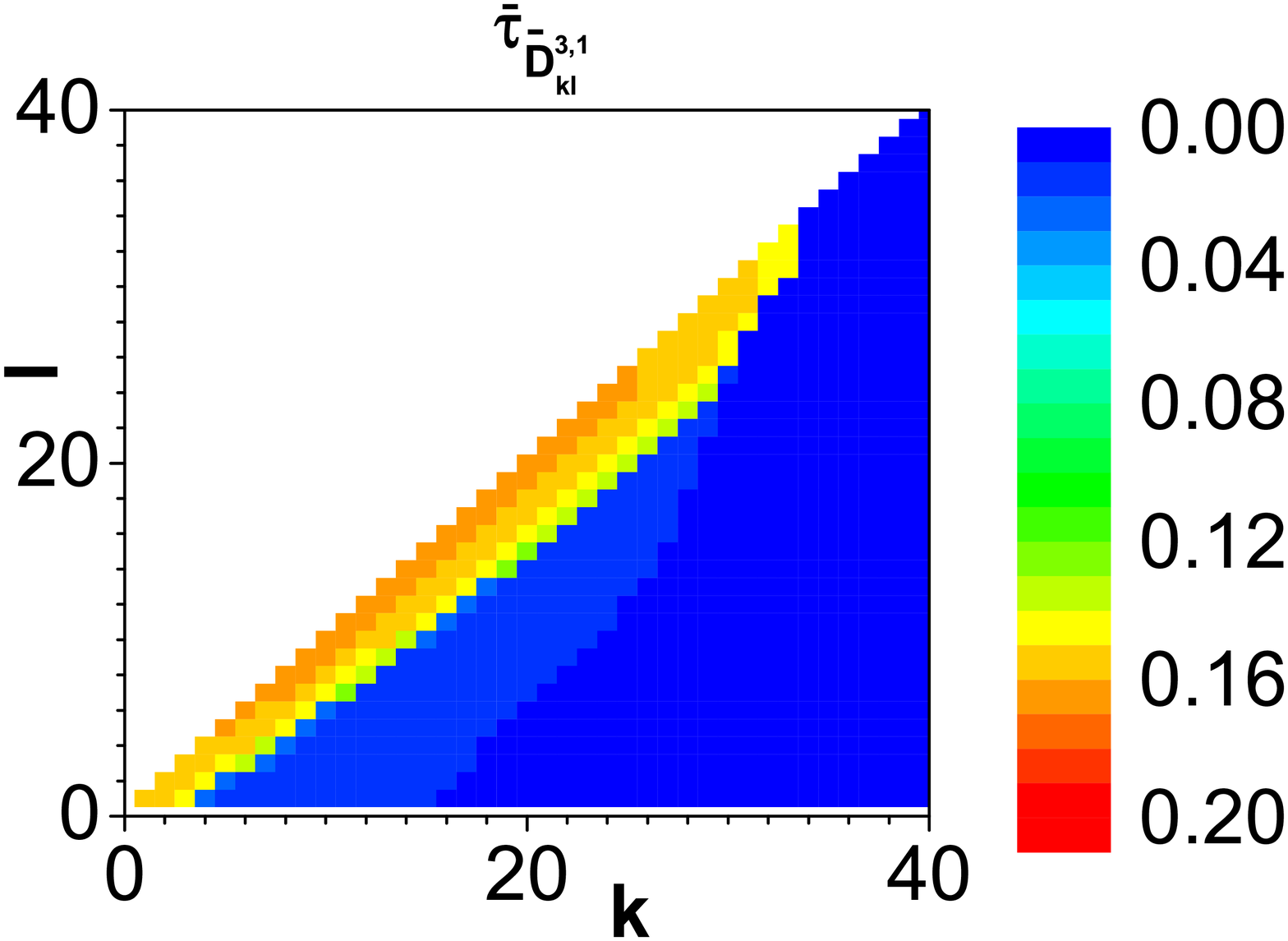}} \\
 \centerline{\small (a) \hspace{.4\hsize} (b)}
 \caption{Non-classicality depth $ \bar{\tau} $ of NCa $ \bar{\cal D}_{kl}^{3,1} $ in Eq.~(\ref{31}) as it depends on
  indices $ k $ and $ l $ for photon-number distribution reconstructed by (a) the maximum-likelihood approach and (b) the Gaussian fit;
  $ M_n = 80 $. Relative experimental errors are lower than 6~\%.}
 \label{fig8}
\end{figure}

We demonstrate the complementarity of the NCa $ ^{\rm s}\bar{\cal D}_{kl}^{2,1}
$ and the NCa $ \bar{\cal D}_{kl}^{3,1} $ using the analyzed twin beam.
According to the graphs in Fig.~\ref{fig7}, the NCa $ ^{\rm s}\bar{\cal
D}_{kl}^{2,1} $ gives the best performance for maximally different values of
the indices $ k $ and $ l $, where we have $ \bar{\tau} \approx 0.25 $.
Contrary to this, the NCa $ \bar{\cal D}_{kl}^{3,1} $ works only for close
values of the indices $ k $ and $ l $, as shown in Fig.~\ref{fig8}. The reached
values of the NCDs are also smaller ($ \bar{\tau} \approx 0.15 $) for $
\bar{\cal D}_{kl}^{3,1} $.

Sensitivity of the NCa to the 'local' noise is illustrated by two examples. In
Figs.~\ref{fig7}(a,b) we compare the NCDs $ \bar{\tau} $ of the NCa $ ^{\rm
s}\bar{\cal D}_{kl}^{2,1} $ for the photon-number distributions of the Gaussian
fit and ideal twin beam. Though the small amount of the noise present in the
Gaussian fit only slightly lowers the maximal attained values of the NCD $
\bar{\tau} $ compared to that for the ideal twin beam, it qualitatively
broadens the area of indices $ k $ and $ l $ where the non-classicality is
observed. In the second example, we compare the performance of the NCa $
\bar{\cal D}_{kl}^{3,1} $ by analyzing two reconstructed photon-number
distributions considerably differing in the amount of the noise. The
corresponding NCDs $ \bar{\tau} $ are plotted in Figs.~\ref{fig8}(a,b). Local
irregularities in the graph of the NCDs $ \bar{\tau} $ belonging to the
photon-number distribution from the maximum-likelihood approach [see
Fig.~\ref{fig8}(a)] compared to that for the Gaussian fit [see
Fig.~\ref{fig8}(b)] are apparent.

\begin{figure} % figs. 9a, b
 \resizebox{.47\hsize}{!}{\includegraphics{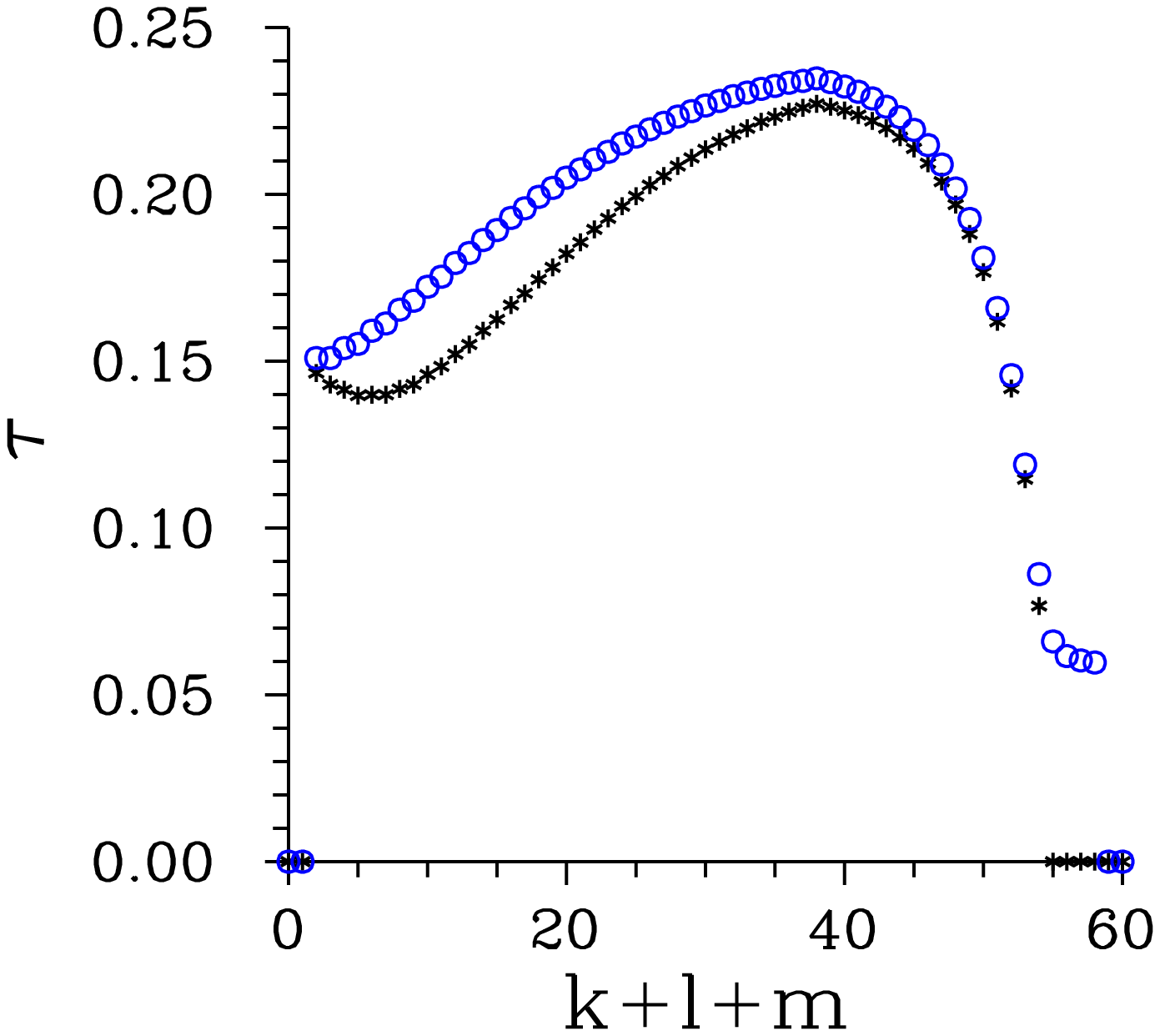}}  \hspace{1mm}
 \resizebox{.47\hsize}{!}{\includegraphics{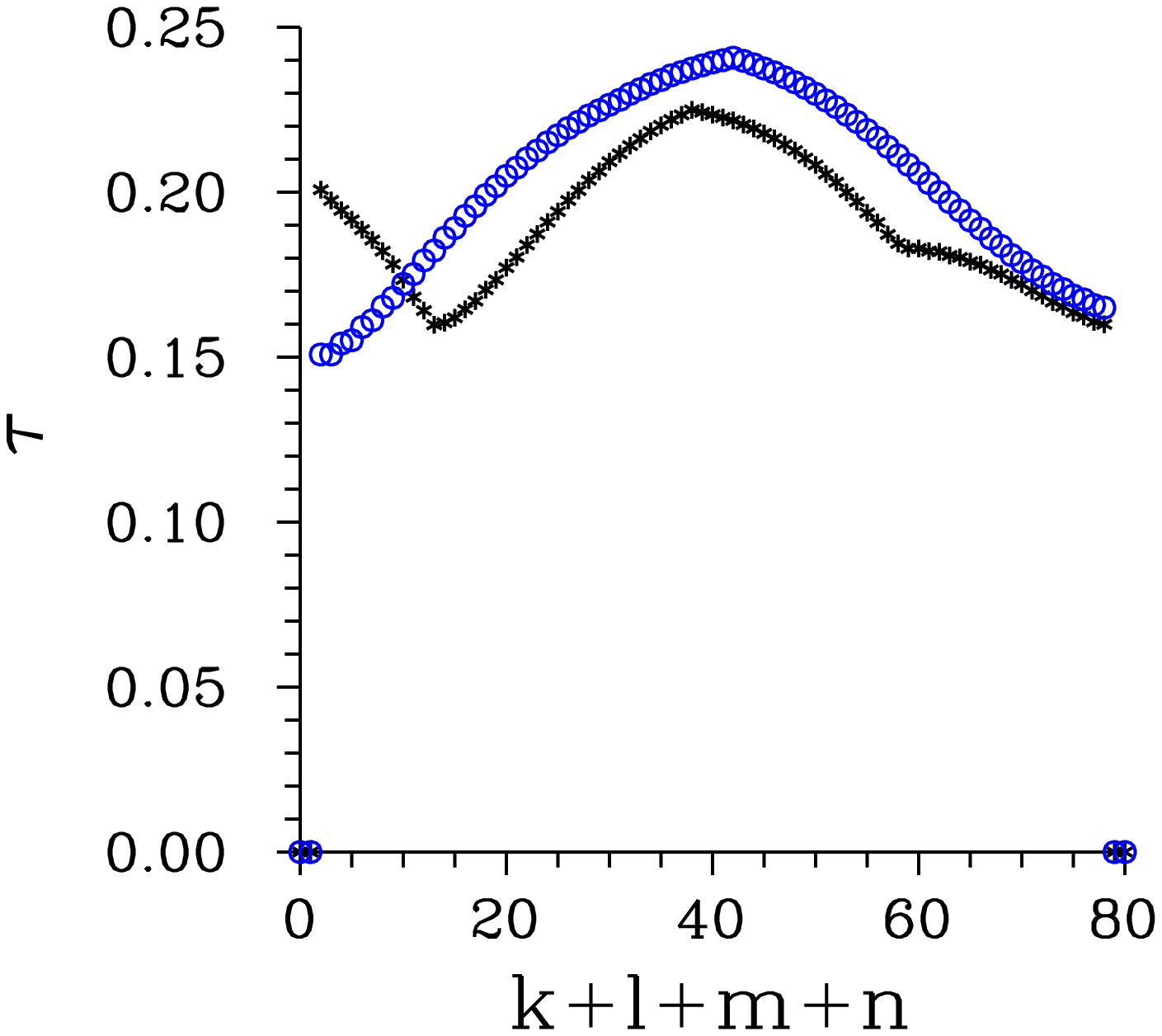}} \\
 \centerline{\small (a) \hspace{.4\hsize} (b)}
 \caption{(a) [(b)] Maximal non-classicality depth $ \bar{\tau} $ of NCa $ ^{\rm s}\bar{\cal D}_{klm} $ [$ \bar{\cal D}_{klmn} $]
  in Eq.~(\ref{26}) [(\ref{32})] as it depends on the sum $ k+l+m $ [$ k+l+m+n $] of indices for photon-number distribution
  reconstructed by the maximum-likelihood approach ($ \ast $) and the Gaussian fit (blue $ \circ $); $ M_n=80 $.
  Relative experimental errors are lower than 6~\%.}
 \label{fig9}
\end{figure}

In the third scenario, we look for the maximal value of the NCD $ \bar{\tau} $
considering all NCa included in a given parametric system. We may monitor this
search, e.g., by sorting the maximal attainable values of the NCD $ \bar{\tau}
$ with respect to the sum of the involved indices. This then allows us to
determine an average value of the index (photon number) corresponding to the
maximal value of $ \bar{\tau} $. As an example, we apply this approach to the
NCa $ ^{\rm s}\bar{\cal D}_{klm} $ in Eq.~(\ref{26}) and $ \bar{\cal D}_{klmn}
$ in Eq.~(\ref{32}) derived from the majorization theory and labelled with
three and four indices, respectively. The corresponding maximal NCDs $
\bar{\tau} $ are plotted in Figs.~\ref{fig9}(a,b) for two reconstructed
photon-number distributions. All curves in Fig.~\ref{fig9} have clearly defined
maxima that provide the average index - photon number around 11-12. We note
that the mean reconstructed photon numbers lie around 9.

In the end, we note that the NCCPs $ \bar{\nu} $ defined in Eq.~(\ref{41}) can
be applied instead of the NCDs $ \bar{\tau} $. They give similar results. On
one hand, they have a greater dynamical range, on the other hand, their values
have no upper bound. Both non-classicality quantifiers depend on the number $ M
$ of considered modes. For the analyzed twin beams, if the mode number $ M $ is
chosen considerably smaller than the actual one ($ M_n $), the NCDs $
\bar{\tau} $ may exceed the value 0.5, which is the expected upper bound for a
single-mode Gaussian state. In case of the NCCPs $ \bar{\nu} $, even their
large values do not have to suffice to completely conceal the twin-beam
non-classicality.

We also note that, in the above analysis of the non-classicality of the
experimental twin beam, we have paid our attention to only those parametric
systems of the NCa for probabilities presented in Sec.~II that suit the best to
the analyzed twin beam. Also other parametric systems of Sec.~II indicate the
non-classicality of the twin beam, though weaker than the used ones. We would
like to stress here that the parametric systems of the NCa of Sec.~II are
derived for arbitrary bipartite optical fields. Other NCa than those applied to
the twin beam may naturally be more suitable for other types of nonclassical
fields.

\section{Conclusions}

Using the Mandel detection formula, we have formulated several parametric
systems of non-classicality criteria for probabilities in bipartite optical
fields. They arose from the mutual comparison of the non-classicality criteria
derived by using nonnegative polynomials in intensities, the Cauchy--Schwarz
inequality, the matrix approach and the majorization theory. This comparison
that revealed relations among various non-classicality criteria allowed us to
identify the fundamental criteria.

We have extended the concept of the Lee non-classicality depth such that it can
be applied directly to photocount and photon-number distributions. We have
pointed out that the number of field modes in this concept is an important
parameter, that is not, however, usually taken into account. This may result in
unexpectedly high values of the non-classicality depth. The non-classicality
counting parameter, that is easily determined, has been revealed as an
alternative quantifier of the non-classicality of the states with weaker forms
of the non-classicality.

The criteria for probabilities can be grouped according to their structure, in
close analogy to the criteria for intensity moments. It has been shown that the
criteria for probabilities and intensity moments from the corresponding groups
give comparable values of the non-classicality quantifiers in the case of twin
beams. The criteria in probabilities show a tendency to give slightly larger
values of the non-classicality quantifiers.

The performance of the derived parametric systems of the non-classicality
criteria has been demonstrated considering three photon-number distributions of
twin beams with different level of the noise. The parametric systems giving the
maximal values of the non-classicality depth in case of twin beams have been
identified. These parametric systems have allowed to visualize the local
distribution of the non-classicality across the photon-number distribution as
well as to make visible the pairwise structure of the twin beam. High
sensitivity of these systems to the level of the noise originating in the
experiment and reconstruction method has been demonstrated.

Though the analysis has been performed for twin beams, the derived and
discussed non-classicality criteria are applicable to any non-classical
bipartite multi-mode optical field.

\acknowledgements \noindent The authors thank GA \v{C}R projects No.
18-08874S(V.M.,O.H.) and No.~18-22102S (J.P.). They also acknowledge the
support from M\v{S}MT \v{C}R (project No.~CZ.1.05/2.1.00/19.0377). The authors
thank M. Hamar for his help with the experiment.

\appendix
\section{Non-classicality criteria for probabilities for low photon (photocount) numbers.}

In this Appendix, we explicitly write the NCa for probabilities for
photon-numbers up to five needed for the analysis of non-classicality of weak
optical fields containing on average up to 2-3 photons. Such fields are met in
various areas of quantum optics. Twin beams exploited in the heralded
single-photon sources may serve as an example
\cite{PerinaJr2001,Alibard2005,Brida2012}. In writing the NCa we keep the
nomenclature developed in ~\cite{PerinaJr2017a} for the NCa based on intensity
moments.

The NCa for intensity moments written in Eq.~(\ref{4}) and derived by
considering nonnegative polynomials of intensities give us the following NCa
for probabilities [$ p(n_{\rm s},n_{\rm i}) \equiv p_{\rm si}(n_{\rm s},n_{\rm
i}) $]:
\begin{eqnarray}  % A1-A13
 &\bar{E}_{001} = p(2,0) + p(0,2) - p(1,1) <0,& \label{A1} \\
 &\bar{E}_{101} = 3p(3,0) + p(1,2) - 2p(2,1) <0,& \label{A2} \\
 &\bar{E}_{011} = 3p(0,3) + p(2,1) - 2p(1,2) <0,& \label{A3} \\
 &\bar{E}_{201} = 6p(4,0) + p(2,2) - 3p(3,1) <0,& \label{A4} \\
 &\bar{E}_{021} = 6p(0,4) + p(2,2) - 3p(1,3) <0,& \label{A5} \\
 &\bar{E}_{111} = 3p(3,1) + 3p(1,3) - 4p(2,2) <0,& \label{A6} \\
 &\bar{E}_{301} = 10p(5,0) + p(3,2) - 4p(4,1) <0,& \label{A7} \\
 &\bar{E}_{031} = 10p(0,5) + p(2,3) - 4p(1,4) <0,& \label{A8} \\
 &\bar{E}_{211} = 2p(4,1) + p(2,3) - 2p(3,2) <0,& \label{A9} \\
 &\bar{E}_{121} = 2p(1,4) + p(3,2) - 2p(2,3) <0,& \label{A10} \\
 &\bar{E}_{002} = p(4,0) + p(2,2) + p(0,4) - p(3,1) - p(1,3)& \nonumber \\
   &<0,& \label{A11} \\
 &\bar{E}_{102} = 5p(5,0) + 3p(3,2) + p(1,4) - 4p(4,1) - 2p(2,3)& \nonumber \\
   & <0,& \label{A12} \\
 &\bar{E}_{012} =  5p(0,5) + 3p(2,3) + p(4,1) - 4p(1,4) - 2p(3,2) & \nonumber \\
   & <0.& \label{A13}
\end{eqnarray}
They correspond to the NCa for intensity moments written in Eqs.~(20)---(32) in
Ref.~\cite{PerinaJr2017a}.

Similarly, the NCa for intensity moments written in Eq.~(\ref{5}) provide us
the following NCa for probabilities [$ p_{\rm s}(n,m) \equiv p_{\rm si}(n,m) $,
$ p_{\rm i}(n,m) \equiv p_{\rm si}(m,n) $]:
\begin{eqnarray}  % A14-A16
 &\bar{E}_{1011} = ^{\rm s}\bar{E}_{1011} <0,& \label{A14} \\
 &\bar{E}_{0111} = ^{\rm i}\bar{E}_{1011} <0,& \label{A15} \\
 & ^{a}\bar{E}_{1011} = 12p_a(3,2)p^4(0,0) + 2[ p_a^2(1,0)p_a(1,2)& \nonumber \\
 & + 3p_a^2(0,1)p_a(3,0) + 4p(0,1)p(1,0)p_a(2,1)] p^2(0,0) & \nonumber \\
 & + p_a^2(0,1)p_a^3(1,0) - 4[2p_a(1,0)p(2,2) + 3p_a(0,1)p_a(3,1)] & \nonumber \\
 & \times p^3(0,0)  - 2[2p_a^2(0,1)p_a(1,0)p_a(2,0) & \nonumber \\
 & + p_a^2(1,0)p_a(0,1)p(1,1)]p(0,0) <0, & \nonumber \\
 & \bar{E}_{0011} = 4p(2,2)p^3(0,0) + 2[\sum_b p_b^2(1,0)p_b(0,2)& \nonumber \\
 & + 2p(1,0)p(0,1)p(1,1)]p(0,0)- 4\sum_b p_b(1,0)p_b(1,2) & \nonumber \\
 & \times p^2(0,0) - 3p^2(1,0)p^2(0,1) <0;& \label{A16}
\end{eqnarray}
symbol $ \sum_b $ denotes summation over indices $ \rm s $ and $ \rm i $. These
NCa are counterparts of the NCa for intensity moments written in Eqs.~(65) and
(66) of Ref.~\cite{PerinaJr2017a}.

The following NCa for probabilities given in Eq.~(\ref{12}) and originating in
the Cauchy--Schwarz inequality are useful for low-intensity fields:
\begin{eqnarray}  % A17-A18
 &\bar{C}_{12}^{10} = 2p(1,2)p(1,0)- p^2(1,1) <0,& \label{A17} \\
 &\bar{C}_{01}^{21} = 2p(2,1)p(0,1)- p^2(1,1) <0.& \label{A18}
\end{eqnarray}
They correspond to the NCa for intensity moments written in Eqs.~(80) and (82)
of Ref.~\cite{PerinaJr2017a}.

The matrix approach based on $ 2\times 2 $ and $ 3\times 3 $ matrices of
Eqs.~(\ref{13}) and (\ref{14}) leaves us with the following NCa for
probabilities:
\begin{eqnarray}  % A19-A21
 &\bar{M}_{1100} = 4p(2,2)p(0,0)- p^2(1,1) <0,& \label{A19} \\
 &\bar{M}_{1001} = 4p(2,0)p(0,2)- p^2(1,1) <0,& \label{A20} \\
 &\bar{M}_{001001} = 4p(2,0)p(0,2)p(0,0) +2p(1,1)p(1,0)p(0,1) & \nonumber \\
 & - p^2(1,1)p(0,0) - 2\sum_b p_b(2,0)p_b^2(0,1) <0.& \label{A21}
\end{eqnarray}
The accompanying NCa for intensity moments are expressed in Eqs.~(73), (74) and
(77) of Ref.~\cite{PerinaJr2017a}.

The majorization theory with three variables suggests the following NCa for
probabilities ($ a = {\rm s,i} $):
\begin{eqnarray}  % A22-A23
 &^{a}\bar{D}_{111}^{210} = 2p_a(2,0)p_a(1,0) + \sum_b p_b(2,1)p(0,0) & \nonumber \\
 & + \sum_b p_b(2,0)p_b(0,1) - 3p_a(1,0)p(1,1) <0,& \label{A22} \\
 &^{a}\bar{D}^{220}_{211} = 2p_a^2(2,0) + 2p(2,2)p(0,0) + 2p(2,0)p(0,2) & \nonumber \\
 & - \sum_b p_b(2,1)p_a(1,0) - p_a(2,0)p(1,1) <0. & \label{A23}
\end{eqnarray}
They correspond to the NCa for intensity moments written in Eqs.~(47) and (48)
of Ref.~\cite{PerinaJr2017a}.

Similarly, the majorization theory with four variables leaves us with the
following NCa for probabilities ($ a = {\rm s,i} $):
\begin{eqnarray}  % A24-A32
 &\bar{D}_{1111}^{2110} = \sum_b p_b(2,1)\sum_c p_c(1,0) + \sum_b p_b(2,0)p(1,1) & \nonumber \\
 & - 3p^2(1,1) <0,& \label{A24} \\
 &\bar{D}_{1111}^{2200} = 2\bigl[\sum_b p_b(2,0)\bigr]^2 + 4p(2,2)p(0,0) - 3p^2(1,1) & \nonumber \\
 & <0,& \label{A25} \\
 &\bar{D}_{1111}^{4000} = 12\sum_b p_b(4,0)p(0,0) - p^2(1,1)<0,& \label{A26} \\
 &^{a}\bar{T}_{1110}^{2100} = 2\sum_b p_b(2,1)p^2(0,0) + 4[ 3p_a(2,0)p_a(1,0) & \nonumber \\
 & + \sum_b p_b(2,0)p_b(0,1)] p(0,0) - 6p_a(1,0)p(1,1)p(0,0) & \nonumber \\
 & - 3p_a^2(1,0)\sum_b p_b(1,0) <0,& \label{A27} \\
 & \bar{T}_{1110}^{2100} = 2\sum_b p_b(2,1)p^2(0,0) + 2[2\sum_b p_b(2,0)p_b(1,0) & \nonumber \\
 & + 3\sum_b p_b(2,0)p_b(0,1)] p(0,0) - 3\sum_b p_b(1,0)p(1,1)p(0,0) & \nonumber \\
 & - 3\sum_b p_b^2(1,0)p_b(0,1) <0,& \label{A28} \\
 & ^{a}\bar{T}_{2110}^{2200} = 4p(2,2)p^2(0,0) + 4[3p_a^2(2,0) + 2p(2,0)p(0,2)] & \nonumber \\
 & \times p(0,0) - 2[\sum_b p_b(2,1)p_a(1,0) + p_a(2,0)p(1,1)] p(0,0) & \nonumber \\
 & - 2p_a^2(1,0)p_a(2,0) - \sum_b p_b(2,0)p_a^2(1,0) & \nonumber \\
 & - 2p_a(2,0)p(0,1)p(1,0) <0,& \label{A29}  \\
 & \bar{T}_{2110}^{2200} = 4p(2,2)p^2(0,0) + 4[\sum_b p_b^2(2,0) + 3p(2,0) & \nonumber \\
 & \times p(0,2)]p(0,0) - [\sum_b p_b(1,0) \sum_c p_c(2,1)  & \nonumber \\
 & + \sum_b p_b(2,0) p(1,1)] p(0,0) - 2 \sum_b p_b(2,0) p(1,0)p(0,1) & \nonumber \\
 & - \sum_b p_b(2,0)p_b^2(0,1)  <0,& \label{A30} \\
 & ^{a}\bar{T}_{1111}^{2110} = 2[\sum_b p_b(2,1)p_a(1,0)+p_a(2,0)p(1,1)]p(0,0)  & \nonumber \\
 &  + [3p_a(2,0) + p_a(0,2)]p_a^2(1,0) + 2 p_a(2,0)p(1,0)p(0,1) & \nonumber \\
 &  - 6p_a^2(1,0)p(1,1) <0,& \label{A31} \\
 & \bar{T}_{1111}^{2110} = [\sum_b p_b(1,0)\sum_c p_c(2,1) +  \sum_b p_b(2,0)p(1,1)] & \nonumber \\
 & \times p(0,0) + 2\sum_b p_b(2,0)p(1,0)p(0,1) & \nonumber \\
 & \hspace{-6mm} + \sum_b p_b(2,0)p_b^2(0,1) -6p(1,0)p(0,1)p(1,1) <0.& \label{A32}
\end{eqnarray}
The corresponding NCa for intensity moments are given in Eqs.~(49)---(55), (75)
and (76) of Ref.~\cite{PerinaJr2017a}.

%\bibliography{perina}

%merlin.mbs apsrev4-1.bst 2010-07-25 4.21a (PWD, AO, DPC) hacked
%Control: key (0)
%Control: author (0) dotless jnrlst
%Control: editor formatted (1) identically to author
%Control: production of article title (0) allowed
%Control: page (1) range
%Control: year (0) verbatim
%Control: production of eprint (0) enabled
%

\end{document}